\documentclass[nofootinbib, amsmath,amssymb, aps,]{revtex4}

\usepackage{hyperref}

\usepackage{graphicx}
\usepackage{dcolumn}
\usepackage{bm}

\usepackage{booktabs}
\usepackage{topcapt}

\usepackage{xcolor}

\usepackage{slashed} 

 \usepackage{axodraw2}
 
 \usepackage{ulem}
 \usepackage{cancel}

\newcommand{\MyInt}{\mathcal{I}_0 ~}
\newcommand{\MyIntPhoton}{\mathcal{I}_2 ~}

\begin{document}
\title{On the Photon-Fermion Vertex}
\author{Orlando Oliveira}

\affiliation{CFisUC, Departament of Physics, University of Coimbra, 3004-516 Coimbra, Portugal}
\affiliation{Instituto Tecnol\'ogico de Aeron\'autica,  DCTA, 12228-900 S\~ao Jos\'e dos Campos,~Brazil}

\begin{abstract}
The QED Dyson-Schwinger equation for the photon-fermion one-particle irreducible Green function, the photon-fermion vertex,
is investigated using 
a Ball-Chiu description for its longitudinal part, together with
the K{\i}z{\i}lersu-Reenders-Pennington basis for its transverse part.
Exact expressions for all the transverse form factors are derived from the vertex Dyson-Schwinger equation.
Furthermore, feeding the Dyson-Schwinger equation with a simplified vertex 
that
goes beyond the perturbative solution, some of the results of the one-loop perturbative  
calculation are recovered in a more general framework.  
The approach allows also exact results for the on-shell vertex, that encode the anomalous magnetic and electric fermion couplings,
and for its soft photon limit. 
The investigation of the chiral limit of the photon-fermion vertex shows that a limited number of transverse form factors are required,
a result that, once more, is in good agreement with a one-loop calculation for QED but that appears in a more general framework. 
The results derived for the photon-fermion vertex can be extended easily to the quark-gluon vertex after proper modifications.
\end{abstract}

\maketitle
\tableofcontents

\section{Introduction and Motivation}

The interaction of photons with fermions, being it leptons, quarks or any composite  charged particle, is described by quantum electrodynamics (QED),
that provides the theoretical framework to explain all sub-atomic phenomena, with the exception of nuclear and sub-nuclear physics. 
The diversity of the observed effects at the level of Atomic and Condensed Matter Physics or Quantum Optics illustrates well the richness of the phenomena
driven by QED.
Furthermore, Quantum Electrodynamics is a gauge theory, associated with the gauge group U(1),
and it guided the building of the non-Abelian gauge theories that have a major role in contemporary Particle Physics. Indeed,
the electroweak theory and quantum chromodynamics (QCD), the color interaction that rules the hadronic world, are examples of non-Abelian gauge theories.

The usual textbook approach to solve QED is perturbation theory. If perturbation theory has a major  role in our understanding 
of the quantum world, it is not the only approach to Quantum Field Theory (QFT). Moreover, 
there are phenomena, such as chiral symmetry breaking  \cite{Ding:2022ows,Binosi:2022djx,Papavassiliou:2022wrb,Alkofer:2023lrl}  that, in QCD, is responsible 
for the generation of mass of the  Universe, that can not be explained within a perturbative framework. Chiral symmetry breaking can also occur in QED
\cite{Miransky:1984ef,Miransky:1985wzx,Gusynin:1986fu,Miransky:1989qc,Bardeen:1990im,Braun:2010qs,Antipin:2012kc}
but for larger values of the coupling constants $\alpha > \pi/3$, to be compared with $\alpha = 1/137$ the standard value for the coupling constant in QED.

Our goal is to explore the Dyson-Schwinger equations (DSE) for QED in a general linear covariant gauge, 
that is defined by adding the term $(\partial A)^2/2 \xi$ to the classical Lagrangian density. In particular we will explore 
the Dyson-Schwinger  equation for the photon-fermion one-particle irreducible Green function. 
In the following the one-particle irreducible Green functions will be named also as vertices.
The DSE are an infinite tower of integral equations that relate all QED Green functions. Then, to solve the DSE it is mandatory to 
introduce a truncation and/or model some vertices, i.e. only a subset of the Green functions will be accessed each time the DSE are solved. 
How far does one has to go in the tower of equations depends on the problem at hand. 

Of the multiple QED Green functions our focus is the photon-fermion vertex and how the corresponding DSE can help in its description. 
The photon-fermion vertex has to comply with the symmetries of QED and with the general principles of QFT as described in e.g.
\cite{Roberts:1994dr,RichardWilliams2007}. This vertex call for twelve form factors and, in momentum space, it is common to 
separate the vertex into a longitudinal and a transverse part, relative to the photon momentum. The longitudinal part of the vertex is described
by four form factors, while the transverse part requires the other eight form factors. Gauge symmetry is expressed through the vertex Wark-Takahashi identity 
(WTI), whose solution writes the four longitudinal form factors in terms of the fermion propagator functions. The WTI identities do not provide a unique
solution to the longitudinal form factors as it is always possible to add a transverse component that does not contribute to WTI. However, herein
we will take the ``minimal solution'' of the WTI and use the simplest longitudinal form factors that solve this identity.
The transverse form factors are not constraint by gauge symmetry and, in practice, they have to be modelled in a way that the DSE
solution complies with the symmetries of QED and multiplicative renormalizability of the theory, i.e. they are modelled to reproduce
the perturbative solution at large momentum. 
In the literature there are various models for the transverse vertex that  conform with multiplicative renormalization, 
at least in an approximate version. Our primary goal is to derive expressions for the transverse part that depend only on the 
Dyson-Schwinger equation for the photon-fermion vertex. By considering a tensor basis for the photon-fermion vertex it is possible to
derive exact kinematical relations, that come from exploring the tensor properties of the basis, and that determine all the transverse form factors.
These exact expressions are valid beyond QED and also apply to QCD, after taking into account the color structure of the interaction. The dynamical
information enters when we try to solve these relations and have to call for the right hand side of the vertex DSE that are not the same 
for QED and for QCD. The handling of the dynamical information in the vertex DSE is a complex task that we solve in QED by approximate the vertices
appearing in the r.h.s. of the equation. 
Moreover, although the formal exact expressions derived for the transverse form factors are worked out in Minkowski spacetime, they are 
scalar equations that can be easily translate into Euclidean spacetime by applying the usual mapping rules to perform the Wick rotation.

The relevance of investigating the QED photon-fermion vertex goes beyond the solution of theory itself as it also helps understanding and
modelling the  QCD quark-gluon vertex and, therefore, our understanding of non-Abelian gauge theories, see e.g.
 \cite{Alkofer:2000wg,Fischer:2006ub,Ferreira:2023fva,Aguilar:2023mdv} for references for QCD. 
For hadronic physics it is crucial to have a proper description of the quark-gluon vertex. 
In a non-Abelian gauge theory the number of fundamental vertices, i.e. those appearing in its Lagrangian density,
is larger than for QED, which makes the DSE for the non-Abelian theories harder to solve. 
For QCD there have been some tentative solutions of the vertex equation, see \cite{Aguilar:2016lbe,Gao:2021wun} and references therein, 
which necessarily introduce approximations and, as always, also some modelling.

Before exploring the photon-fermion vertex DSE, the fermion and photon propagator equations are also considered. Although no solution
of the equations is discussed, having its full expression in terms of form factors helps understanding the importance of the contribution
of the form factors to the mechanism of mass generation and also their relevance for the fermion wave function. 
This enlarged set of Dyson-Schwinger equations to be used here are well known and have been rederived in \cite{Oliveira:2022bar},
together with the vertex Ward-Takahashi identity (WTI) and the two-photon-two-fermion WTI. Moreover,
in \cite{Oliveira:2022bar} the solution of the two-photon-two-fermion WTI was build and it will be used in the current work. 
We refer the reader to this work for details on the notation and on the definitions used throughout the current manuscript.

In order to help establishing the notation, let us write the photon propagator in a general linear covariant gauge,
defined by  the gauge fixing parameter $\xi$, in Minkowski spacetime
\begin{equation}
    D_{\mu\nu}(k) = - \left( g_{\mu\nu} - \frac{k_\mu k_\nu}{k^2} \right) \, D(k^2) - \frac{\xi}{k^2} \, \frac{k_\mu k_\nu}{k^2} 
    = - P^\perp_{\mu\nu} (k) \, D(k^2) - \frac{\xi}{k^2} \, P^L_{\mu \nu}(k)  \ .
\end{equation}
The Lorentz invariant scalar function $D(k^2)$ will also be referred as the photon propagator. 
The inverse of the fermion propagator is given by
\begin{equation}
 S^{-1}(p) = A(p^2) \, \slashed{p} - B(p^2) + i \, \epsilon \ ,
\end{equation}
where $A(p^2)$ and $B(p^2)$ are Lorentz scalar functions. In general, in the following and unless clearly stated, the $i \epsilon$ term will be 
omitted from now on.

The manuscript is organized as follows. In Sec. \ref{Sec:basico} the basic DSE will be given and will be used to help setting the notation.
In Sec. \ref{Sec:Renormalizacao} the issues related with the renormalization  of QED will be discussed. The tensor basis description
of the photon-fermion vertex is introduced in Sec. \ref{Sec:vertice-tensor} and used to write the propagator equations in Secs. \ref{Sec:PropEqGap}
and \ref{Sec:PropEqPhoton}. A bref analysis of possible Ir divergences is also worked out.
The photon-fermion vertex DSE is studied in detail in Sec. \ref{SecVertex} based on
the basis introduced previously. In Sec. \ref{Sec:rhsVertex} we discuss a model for the photon-fermion vertex that can be used in understanding
and predict the various transverse form factors. This Sec. can be omitted on the first reading. Then, we proceed and discuss the vertex in various
cases. The on-shell vertex and its chiral limit are derived in Sec \ref{Sec:OnShell} and the contribution of the photon-fermion form factors to each of the
on-shell form factors are given explicitly. Corresponding expressions for the chiral limit of the on-shell vertex are also provided.
In \ref{Sec:softphotonlimit} the soft limit of the photon-fermion vertex, that results in integral constraints on the form factors, is discussed.
In \ref{Sec:ChiralVertex} the chiral limit of the photon-fermion vertex is discussed and the results compared with the outcome of one-loop perturbation
theory. In Sec. \ref{SecTransverseFF} the vertex DSE is translated into exact expressions for each of the transverse form factors. These expressions
are derived in Minkowski spacetime but are easily adapted to Euclidean spacetime. Finally, in Sec. \ref{Sec:Summary} we summarise and conclude.

\section{The Dyson-Schwinger equations for QED \label{Sec:basico}}

\begin{figure}[t] 
   \centering
   \includegraphics[width=5.5in]{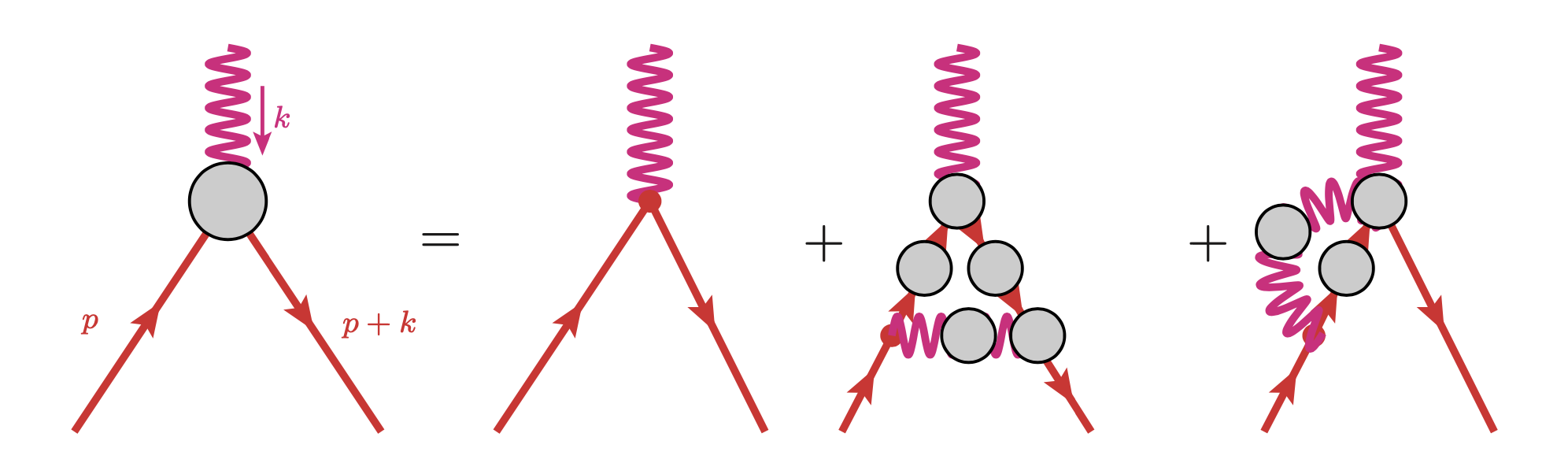} 
\caption{The Dyson-Schwinger equation for the photon-fermion vertex.}
\label{Fig:DSEvertex}
\end{figure}

In this section we provide the basic equations to be studied later and introduction further notation. 
Details on the derivation of the equations and definitions can be found in \cite{Oliveira:2022bar}; see also e.g. \cite{Roberts:1994dr} and references
therein. 
In momentum space, the bare Dyson-Schwinger for the fermion propagator, in Minkowski spacetime, reads
\begin{equation}
  S^{-1} (p)  = \left(  \slashed{p} - m \right)
     - \, i \, g^2 \, \int \frac{d^4 k}{( 2 \, \pi )^4} ~   D_{\mu\nu} (k) ~  \Big[ \gamma^\mu ~ S(p - k) ~   {\Gamma}^\nu (p-k, - p; k)  \Big] \ ,
     \label{DSE-Fermion}
\end{equation}
while the bare equation for the photon propagator in a linear covariant gauge $\xi$ is 
\begin{eqnarray}
  \frac{1}{D(k^2)}  = 
   k^2 
  - i \, \frac{g^2}{3} \,  \int \frac{d^4 p}{(2 \, \pi)^4} ~ \text{Tr} \Big[ \gamma_\mu \, S(p) \, {\Gamma}^\mu(p, -p + k; -k ) \, S(p - k ) \Big] \ .
     \label{Eq:DSE-Photon}
\end{eqnarray}
The bare Dyson-Schwinger equation for the photon-fermion vertex is given by
\begin{eqnarray}
& & 
   {\Gamma}^\mu (p, \, -p -k; \, k)   =  \gamma^\mu  ~ + ~ i \, g^2 \, \int \frac{d^4q}{(2 \, \pi)^4} ~D_{\zeta\zeta^\prime}(q)  ~ \gamma^\zeta ~ S(p-q)
   \nonumber \\
& & 
\qquad
   \Bigg\{ ~ 
                                 \, {\Gamma}^\mu ( p - q, \, -p -k + q; \, k) \,
                                S(p+k-q) \, {\Gamma}^{\zeta^\prime} (p +k-q, \, -p-k; q) 
~ + ~
                                 \, {\Gamma}^{\zeta^\prime\mu} (p - q, \, -p-k; \, q , \, k) 
                                \Bigg\} 
                                \label{Eq:DSE-vertex}
\end{eqnarray}
and requires, besides the propagators and the vertex $\Gamma^\mu$ itself, the one particle irreducible two-photon-two-fermion Green function
$\Gamma^{\mu\nu}$. The diagrammatic representation of the vertex equation is provided in Fig. \ref{Fig:DSEvertex}.
To draw the Feynman diagrams we used \textit{axodraw} \cite{axodraw}.
The two vertices appearing in Eq. (\ref{Eq:DSE-vertex}) satisfy their own WTI and $\Gamma^{\mu\nu}$ is the solution of a DSE that calls
for higher order Green functions. An approximation DSE for this later vertex was derived in \cite{Oliveira:2022bar} but it will not be
considered in this work.

\section{Renormalization in QED \label{Sec:Renormalizacao}}

The Dyson-Schwinger equations written in the previous section are bare versions and to make them finite the theory has to be renormalized. 
Proceeeding as usual, let us introduce the renormalization constants $Z_i$ and defined the physical quantities as
\begin{eqnarray}
& &
   A_\mu  = Z^{\frac{1}{2}}_3 \, A^{(phys)}_\mu   , \quad
   \psi  = Z^{\frac{1}{2}}_2 \, \psi^{(phys)} , \quad
   g = \frac{Z_1}{Z_2 \, Z^{\frac{1}{2}}_3}      g^{(phys)}    ,  
   \nonumber \\
   & & 
   m = \frac{Z_0}{Z_2} \, m^{(phys)} \quad\mbox{and}\quad
  \xi = Z_3 \,  \xi^{(phys)}  .
\end{eqnarray}
For QED the vertex WTI requires $Z_1 = Z_2$, see e.g. \cite{Roberts:1994dr,Oliveira:2022bar}. Then, the renormalized gap equation (\ref{DSE-Fermion}) is given by
\begin{eqnarray}
  S^{-1} (p)  & = &  Z_2 \,   \slashed{p} - Z_0 \, m 
     - \, i \, g^2  \, Z_2 \, \int \frac{d^4 k}{( 2 \, \pi )^4} ~   D_{\mu\nu} (k) ~  \Big[ \gamma^\mu ~ S(p - k) ~   {\Gamma}^\nu (p-k, - p; k)  \Big]    \nonumber  \\
     & = & 
     Z_2 \,   \slashed{p} - Z_0 \, m 
     - \, i \,   g^2 \, Z_2 \, \Sigma(p) \ ,
     \label{DSER-gap}
\end{eqnarray}
the renormalized photon gap equation (\ref{Eq:DSE-Photon}) is
\begin{eqnarray}
  \frac{1}{D(k^2)}  & = &
  Z_3 \,  k^2 
  - i \, \frac{g^2}{3} \,  Z_2 \,  \int \frac{d^4 p}{(2 \, \pi)^4} ~ \text{Tr} \Big[ \gamma_\mu \, S(p) \, {\Gamma}^\mu(p, -p + k; -k ) \, S(p - k ) \Big] \nonumber \\
  & = & Z_3 \,  k^2  \left( 1  - i \,  g^2 \,  \frac{Z_2}{Z_3} \,   \Pi(k^2) \right)
     \label{DSER-photon}
\end{eqnarray}
where $\Sigma (p)$ and $\Pi (k^2)$ are the fermion and the photon physical self-energies, respectively.
The renormalized Dyson-Schwinger equation for the vertex (\ref{Eq:DSE-vertex}) reads
\begin{eqnarray}
   {\Gamma}^\mu (p, \, -p -k; \, k)  &  =  & Z_2 \, \gamma^\mu  ~ + ~ i \, g^2 \,  Z_2 \, \int \frac{d^4q}{(2 \, \pi)^4} ~D_{\zeta\zeta^\prime}(q) \nonumber \\
   & &  \Bigg\{
                                \gamma^\zeta \, S(p-q) \, {\Gamma}^\mu ( p - q, \, -p -k + q; \, k) \,
                                S(p+k-q) \, {\Gamma}^{\zeta^\prime} (p +k-q, \, -p-k; q) 
                                 \nonumber \\
   & & \hspace{7cm} ~ + ~
                                \gamma^\zeta \, S(p-q) \, {\Gamma}^{\zeta^\prime\mu} (p - q, \, -p-k; \, q , \, k) 
                                \Bigg\}  \ .
                                \label{DSER-vertex}
\end{eqnarray}
In the renormalized Eqs (\ref{DSER-gap}) to (\ref{DSER-vertex}) we omitted the index $(ph)$ to simplify the notation. 

The renormalization constants
are fixed by imposing, for example, that
\begin{equation}
A(\mu^2_F) = 1, \qquad B(\mu^2_F) = m \qquad\mbox{ and }\qquad D(\mu^2_B) = \frac{1}{\mu^2_B} 
  \label{Renormalization-Conditions}
\end{equation}
where $\mu_F$ and $\mu_B$ are the renormalization mass scales for the fermion and boson fields, respectively, and $m$ is the physical mass.
For the determination of $Z_0$ and $Z_2$ it is usual to write $\Sigma(p) = \Sigma_v(p^2) \, \slashed{p} + \Sigma_s(p^2)$.
The expressions for $\Sigma_v(p^2)$ and $\Sigma_s(p^2)$ can be read from Eqs (\ref{DSE-Fermion-vector}) and (\ref{DSE-Fermion-scalar}), 
respectively, and from the normalization conditions it comes that
\begin{eqnarray}
    Z_2 &=& {1 \over 1 - i \, g^2 \, \Sigma_v (\mu_F^2)} \, , \\
    Z_0 &=& 1  - i \,  {Z_2 \over m } g^2 \, \Sigma_s (\mu_F^2) \, ,  \\
    Z_3 &=& 1  + i \, Z_2 \, g^2 \, \Pi(\mu_B^2) \, ,
\end{eqnarray}
The combination $g^2 \, D(k^2)$ is independent of the renormalization scale and can be used to define an effective charge for QED.

In perturbation theory the solution of QED requires an IR regulator as the theory is plagued with divergences 
at low momentum, that are associated with the photon being a massless particle. 
For the DSE, possible IR divergences can be look for by studying the renormalized DSE at zero momentum. 
This is better achieved after inserting a tensor basis for the photon-fermion vertex as discussed in Sec. \ref{Sec:vertice-tensor}. 
The analysis of the DSE for the propagators shows no IR divergences.

The IR properties associated with the vertex equation are a bit more difficult to describe, in particular, because this equation requires the
knowledge of the two-photon-two-fermion vertex when one of the photon momenta vanishes. 
In principle the IR divergences appear when the zero photon momentum limit is taken.  
The tensor decomposition of $\Gamma^{\mu\nu}$ calls for a large number of form factors whose IR properties are not known, even in a perturbative 
approach. For the kinematics under consideration, in \cite{Oliveira:2022bar} the WTI for the two-photon-two-fermion vertex was solved and, module possible 
contributions from orthogonal operators relative to the non-vanishing photon momentum, the solution is given in terms of longitudinal
form factors that are functions of $A(p^2)$ and $B(p^2)$ that describe the inverse fermion propagator; see Sec. \ref{Sec:vertice-tensor}
and Eq. (\ref{Eq:Gamma-mu-nu-q0}). If one uses this solution for $\Gamma^{\mu\nu}$ in the vertex equation, then, once more, no IR divergences
seem to be found.

\section{The  photon-fermion vertex \label{Sec:vertice-tensor}}

The description of the photon-fermion vertex $\Gamma^\mu$ calls for twelve form factors \cite{Ball:1980ay}.
It is usual to decomposed  $\Gamma^\mu$ into a longitudinal $\Gamma_L$ and a transverse $\Gamma_T$ part, relative to the photon 
momentum, and write
\begin{equation}
 \Gamma^\mu (p_2, \, p_1; \, p_3)  = \Gamma^\mu_L (p_2, \, p_1; \, p_3) +  \Gamma^\mu_T (p_2, \, p_1; \, p_3) \ ,
\end{equation}
where $p_2$ is the incoming fermion momentum, $-p_1$ is the outgoing fermion momentum and $p_3$ is the incoming photon momentum.
Given that all the momenta are incoming they verify the relation $p_1 + p_2 + p_3 = 0$. From the definition of the transverse 
vertex it follows that
\begin{equation}
 p_{3 \, _\mu } \,  \Gamma^\mu_T (p_2, \, p_1; \, p_3)  = 0 \ .
 \end{equation}
To compute a solution of Eq. (\ref{Eq:DSE-vertex}) it is helpful to introduce a tensor basis and write the longitudinal and transverse vertex components
using the chosen  basis. The two vertex components are then given by
 \begin{eqnarray}
 \Gamma_{L \, \mu} (p_2, \, p_1; p_3) & = & \sum^4_{i=1} \lambda_i (p^2_1, \, p^2_2, \, p^2_3) \, L^{(i)}_\mu (p_1, \, p_2, \,  p_3) \ ,  
 \label{Eq:photon-fermion_vertex-longitudinal}  \\
  \Gamma_{T \, \mu} (p_2, \, p_1; p_3) & = &  \sum^8_{i=1} \tau_i (p^2_1, \, p^2_2, \, p^2_3) \, T^{(i)}_\mu (p_1, \, p_2, \,  p_3) \ ,
 \label{Eq:photon-fermion_vertex-transverse}
\end{eqnarray}
where $L^{(i)}_\mu$ and $T^{(i)}_\mu$ are the set of tensor operators that define the basis for the vertex, and $\lambda_i$ and $\tau_i$
are Lorentz scalar form factors. Note the different ordering of the momenta in the l.h.s. and r.h.s in Eqs. (\ref{Eq:photon-fermion_vertex-longitudinal})
and  (\ref{Eq:photon-fermion_vertex-transverse}). For the longitudinal part of the vertex we take the Ball-Chiu basis \cite{Ball:1980ay} given by 
\begin{eqnarray}
L^{(1)}_\mu (p_1, \, p_2, \,  p_3) & = & \gamma_\mu \ ,  \label{TensorBasis-L1} \\
L^{(2)}_\mu (p_1, \, p_2, \,  p_3) & = & \big( \slashed{p}_1 - \slashed{p}_2 \big) \big( p_{1} - p_{2} \big)_\mu \ , \\
L^{(3)}_\mu (p_1, \, p_2, \,  p_3) & = & \big( p_{1} - p_{2} \big)_\mu \ , \\
L^{(4)}_\mu (p_1, \, p_2, \,  p_3) & = & \sigma_{\mu\nu} \big( p_{1} - p_{2} \big)^\nu \ ,
\label{TensorBasis-Long} 
\end{eqnarray}
while for the orthogonal part of the vertex we rely on the K{\i}z{\i}lersu-Reenders-Pennington basis \cite{Kizilersu:1995iz} 
\begin{eqnarray}
T^{(1)}_\mu (p_1, \, p_2, \,  p_3) & = & p_{1 \, _\mu} \big( p_2 \cdot p_3 \big) - p_{2 \, _\mu}  \big( p_1 \cdot p_3 \big)  \ ,\\
T^{(2)}_\mu (p_1, \, p_2, \,  p_3) & = & - \, T^{(1)}_\mu (p_1, \, p_2, \,  p_3) ~ \big( \slashed{p}_1 - \slashed{p}_2 \big)  \ , \\
T^{(3)}_\mu (p_1, \, p_2, \,  p_3) & = & p^2_3 \, \gamma_\mu - p_{3 \, _\mu} \, \slashed{p}_3  \ , \\
T^{(4)}_\mu (p_1, \, p_2, \,  p_3) & = & T^{(1)}_\mu (p_1, \, p_2, \,  p_3) ~ \sigma_{\alpha\beta} \, p^\alpha_1 \,  p^\beta_2 \ , \\
T^{(5)}_\mu (p_1, \, p_2, \,  p_3) & = & \sigma_{\mu\nu} \, p^\nu_3 \ , \\
T^{(6)}_\mu (p_1, \, p_2, \,  p_3) & = & \gamma_\mu \big( p^2_1 - p^2_2 \big) + \big( p_{1} - p_{2} \big)_\mu \, \slashed{p}_3 \ , \\
T^{(7)}_\mu (p_1, \, p_2, \,  p_3) & = &  - \, \frac{1}{2} \, \big( p^2_1 - p^2_2 \big) \, \big[ \gamma_\mu \,  \big( \slashed{p}_1 - \slashed{p}_2 \big)  - \big( p_{1} - p_{2} \big)_\mu\big] 
               - \big( p_{1} - p_{2} \big)_\mu ~ \sigma_{\alpha\beta} \, p^\alpha_1 \,  p^\beta_2 \ , \\
T^{(8)}_\mu (p_1, \, p_2, \,  p_3) & = &  - \, \gamma_\mu \, \sigma_{\alpha\beta} \, p^\alpha_1 \,  p^\beta_2  \, + \, p_{1 \, _\mu} \slashed{p}_2 \, - \,  p_{2 \, _\mu} \slashed{p}_1  \ ,\label{TensorBasis-Ortho} 
\end{eqnarray}
that is free of kinematical singularities. In Eqs (\ref{TensorBasis-L1}) to (\ref{TensorBasis-Ortho}) we take 
$\sigma_{\mu\nu} = \frac{1}{2} \, [ \gamma_\mu \, , \, \gamma_\nu ]$. 

The longitudinal form factors are determined by the Ward-Takahashi vertex identity, see e.g. \cite{Ball:1980ay} and \cite{Oliveira:2022bar}. 
The solution of the WTI is
\begin{eqnarray}
  \lambda_1 (p^2_1, \, p^2_2, \, p^2_3) & = & \frac{1}{2} \bigg( A\big( p^2_1 \big)  + A\big(p^2_2\big) \bigg)  \ ,    \label{EQ:L1} \\
  \lambda_2 (p^2_1, \, p^2_2, \, p^2_3) & = & \frac{1}{2 \, \left( p^2_1 - p^2_2 \right)}   \bigg( A\big( p^2_1 \big)  -  A\big(p^2_2\big) \bigg) \ , \label{EQ:L2} \\
  \lambda_3 (p^2_1, \, p^2_2, \, p^2_3) & = & \frac{1}{ p^2_1 - p^2_2 }   \bigg( B\big(p^2_1\big) - B\big( p^2_2 \big)    \bigg) \ ,   \label{EQ:L3} \\
  \lambda_4 (p^2_1, \, p^2_2, \, p^2_3) & = & 0   \label{EQ:L4}
\end{eqnarray}
that writes the longitudinal form factors $\lambda_i$ in terms of the fermion propagators functions $A$ and $B$.
If $A$ and $B$ are smooth functions, then $\lambda_2$ and $\lambda_3$ are regular in the limit of $p^2_1 \rightarrow p^2_2$
and become proportional to the derivatives of $A$ and $B$, respectively, at $p^2_1 = p^2_2$. A straightforward calculation shows that
\begin{equation}
  \lambda_1 (p^2, \, p^2, 0) = A(p^2) \ , \qquad
  \lambda_2 (p^2, \, p^2, 0) = \frac{1}{2} \,  \frac{d A(p^2)}{d p^2} \qquad\mbox{ and }\qquad
  \lambda_3 (p^2, \, p^2, 0) = \frac{d B(p^2)}{d p^2} \ .
  \label{LongVertex-ZeroMom}
\end{equation}
that comply with the Ward identity
\begin{equation}
 \Gamma^\mu(p, \, p; \, 0) = \frac{\partial S^{-1} (p)}{\partial p_\mu}  \ .
\end{equation}
For QCD, the Ward-Takahashi identity is replaced by a Slavnov-Taylor identity whose solution for the $\lambda_i$ was worked out in 
\cite{Aguilar:2010cn}. It turns out that for QCD in the soft gluon limit,
defined for a vanishing gluon momenta, and assuming a smooth behaviour $\lambda_2$ and $\lambda_3$ are
 given by derivatives of $A$ and $B$ as described in \cite{Oliveira:2018fkj}.

In the limit of vanishing photon momenta (soft photon limit), the vertex is entirely determined by the longitudinal form factors
$\lambda_i$ as the transverse components of  the vertex do not contribute, as long as they do not have kinematic singularities.
In this case the vertex is described only by $\Gamma_L^\mu$ with the $\lambda_i$  given as in 
Eqs (\ref{LongVertex-ZeroMom}) and (\ref{EQ:L4}).

Before proceeding with the analysis of the vertex DSE let us comment on the available lattice simulations. Lattice simulations 
are a first principles approach to QFT and, therefore, the agreement with their resutls is important to our understanding of the theory
and also to control the systematics of the various approaches. 
The longitudinal form factors $\lambda_i$ where computed, in the soft gluon limit, using full QCD simulations in \cite{Kizilersu:2021jen}. 
On the other hand, in \cite{Sternbeck:2019twy} a lattice calculation of the photon-fermion vertex can be found.

\section{The form factor contributions for the Dyson-Schwinger equations \label{Sec:PropEqFF}}

In the previous section we introduced a tensor basis to describe the photon-fermion vertex. Here we aim to
use this basis to write the Dyson-Schwinger equations  in terms of the form factors associated with each of the 
basis elements. Our analysis start by looking at the DSE for the two point correction functions and then proceed with 
the analysis of the photon-fermion vertex itself. For the Dirac algebra we relied on FeynCalc \cite{FeynCalc1,FeynCalc2,FeynCalc3},
a \textit{Mathematica} software package.

\subsection{The fermion gap equation \label{Sec:PropEqGap}}

The fermion gap equation (\ref{DSE-Fermion}) can be projected into a Dirac scalar and a Dirac vector term. 
The scalar component of the fermion gap equation is obtained by taking the Dirac trace of Eq. (\ref{DSE-Fermion}) and, in a general liinear covariant
gauge, is given by
\begin{eqnarray}
& &
  B (p^2)  =  m 
     - \, i \, g^2 \, \int \frac{d^4 k}{( 2 \, \pi )^4} ~  \frac{1}{A^2((p-k)^2) \,(p-k)^2 - B^2((p-k)^2)}  \Bigg\{
     \nonumber \\
     & & 
D(k^2)
 \Bigg( ~~
     A((p-k)^2) 
      \Bigg[ ~ 
      2 \, \lambda_3 \bigg( \frac{(pk)^2}{k^2} -  p^2\bigg)
              ~ + ~ \tau_1 \bigg( k^2 \, p^2 - (pk)^2 \bigg)
      \nonumber \\
      & & \hspace{4cm}
              + ~ \tau_4 \bigg( k^4 \, p^2 - k^2 \, p^2 \,  (pk) - k^2 \, (pk)^2 + (pk)^3 \bigg)
              ~ + ~ 3 \, \tau_5 \bigg( k^2 - (pk) \bigg)
      \nonumber \\
      & & \hspace{4.5cm}
           + ~   \tau_7 \bigg( \frac{3 }{2} \, k^4 + 4 \, k^2 \, p^2 - \frac{15 }{2} \, k^2 \,  (pk) - 6 \, p^2 \, (pk) + 8 \,  (pk)^2 \bigg)
      \Bigg]
      \nonumber \\
      & & \hspace{2.5cm}
     + ~ B((p-k)^2)  
           \Bigg[
              3 \, \lambda_1
              ~ + ~ 4 \, \lambda_2 \bigg( p^2 - \frac{ (pk)^2}{k^2} \bigg)
              ~ + ~ 2 \, \tau_2 \bigg( k^2 p^2 -  (pk)^2\bigg)
              \nonumber \\
              & & \hspace{8cm}
              ~ + ~ 3 \,  \tau_3 \,  k^2
              + ~ 3 \, \tau_6 \bigg( 2 \, (pk) -  k^2 \bigg)
            \Bigg] ~~
\Bigg)
\nonumber \\
& &
+ ~ \frac{\xi}{k^2} 
    \Bigg(
       A((p-k)^2)  \Bigg[   \lambda_3 \bigg( -\frac{2 (pk)^2}{k^2} + 3 \, (pk) - k^2 \bigg)
                         \Bigg]
       +B((p-k)^2) \Bigg[ \lambda_2 \left(\frac{4 (pk)^2}{k^2} - 4 \, (pk) + k^2\right) +  \lambda_1 \Bigg]
    \Bigg)
    ~\Bigg\}     \ .
      \label{DSE-Fermion-scalar}
\end{eqnarray}
On the other hand, the vector component of the gap equation is computed after multiplying Eq. (\ref{DSE-Fermion})
by $\slashed{p}$ and then taking the Dirac trace. The equation is
\begin{eqnarray}
& & 
p^2 \, A(p^2)  = p^2 
     + \, i \, g^2 \, \int \frac{d^4 k}{( 2 \, \pi )^4} ~   \frac{1}{A^2((p-k)^2) \,(p-k)^2 - B^2((p-k)^2)}\Bigg\{
\nonumber \\
& & 
D(k^2) 
 \Bigg(
     A((p-k)^2) \Bigg[ ~
         \lambda_1 \bigg( - p^2 + 3 \, (pk) - \, 2 \, \frac{(pk)^2}{k^2}\bigg)
        \nonumber \\
        & & \hspace{4cm}
        + ~ 2 \, \lambda_2 \bigg( k^2 \, p^2 + 2 \, p^4 - 2 \, p^2 \, (pk) -  (pk)^2 - \, 2 \, p^2  \, \frac{ (pk)^2}{k^2} + 2 \, \frac{(pk)^3}{k^2} \bigg)
         \nonumber \\
        & &     \hspace{4.5cm}
         + ~ \tau_2 \bigg(k^4 \, p^2 + 2  \, k^2  \, p^4 - 2 \,  k^2  \, p^2  \, (pk) - k^2  \, (pk)^2 - 2 \,  p^2  \, (pk)^2 + 2  \, (pk)^3\bigg)
        \nonumber \\
        & & \hspace{5cm}        
        + ~ \tau_3 \bigg( - \, k^2 \, p^2 + 3 \, k^2 \, (pk) - 2 \, (pk)^2\bigg)
        \nonumber \\
        & & \hspace{5.5cm}        
        + ~ 3 \, \tau_6 \bigg(  k^2 p^2 -  k^2 (pk) - 2 p^2 (pk) + 2 (pk)^2\bigg)
        ~ + ~ 2\, \tau_8 \bigg( k^2 \, p^2 -  (pk)^2\bigg)
                     \Bigg]
     \nonumber \\
     & &  \qquad\qquad
     +~ B((p-k)^2)  \Bigg[~
                              2 \, \lambda_3 \bigg(\frac{(pk)^2}{k^2} -  p^2\bigg)
                             ~ + ~ \tau_1 \bigg(k^2 \, p^2 - (pk)^2\bigg)
                             ~ + ~ \tau_4 \, (pk) \, \bigg(k^2 p^2 -(pk)^2\bigg)
        \nonumber \\
        & & \hspace{4cm}        
                             + ~ 3 \, \tau_5 \, (pk)
                             ~ + ~\tau_7 \bigg(- \, 2 \, k^2 \, p^2 + \frac{3}{2} \,  k^2 (pk) + 6 \, p^2 \, (pk) - 4 (pk)^2\bigg)
                             \Bigg]
     \Bigg)
 \nonumber \\
 & & 
 + ~ \frac{\xi}{k^2}
   \Bigg(
   A(p-k)^2)  \Bigg[
               \lambda_1 \bigg(- p^2 - (pk) + 2 \, \frac{ (pk)^2}{k^2} \bigg)
               \nonumber \\
               & & \hspace{4cm}
               + ~ \lambda_2 \bigg(  p^2 \, k^2 - k^2 \,  (pk)- 4 \, p^2 \, (pk) + 4 \,  (pk)^2 + 4 \, p^2 \,\frac{(pk)^2}{k^2} - 4 \, \frac{(pk)^3}{k^2} \bigg)
               \Bigg]
   \nonumber \\
   & & \qquad\qquad
   +B(p-k)^2) \Bigg[ \lambda_3 \bigg( (pk) - 2 \, \frac{(pk)^2}{k^2}\bigg)
                     \Bigg] ~
   \Bigg)
~ \Bigg\} \ .
     \label{DSE-Fermion-vector}
\end{eqnarray}
In the above two equations the notation
\begin{equation}
\lambda_i = \lambda_i (p^2, \, (p-k)^2,\, k^2 ) \qquad\mbox{ and }\qquad
\tau_i = \tau_i ( p^2, \, (p-k)^2,\, k^2 ) 
\end{equation}
was used to simplify the writing. This decomposition of the fermion gap equation in terms of vertex form factors holds both for QED and for QCD, after
integrating the color degrees of freedom and considering also the contribution of $\lambda_4$ that in QED vanishes.
As can be seen, all the transverse form factors contribute to the scalar and/or vector components of the
gap equation. Note, however, that in the case of $B$ equation only $\tau_8$ does not appear in the r.h.s. of the equation. 

For zero fermion momentum, the scalar gap equation simplifies and gets contributions from a subset of the form factors that appear
 in Eq. (\ref{DSE-Fermion-scalar}). Indeed, for $p = 0$ the equation become
\begin{eqnarray}
  B (0)  & = &  m 
     - \, i \, g^2 \, \int \frac{d^4 k}{( 2 \, \pi )^4} ~  \frac{1}{A^2(k^2) \,k^2 - B^2(k^2)} 
     \nonumber \\
     & & 
      \Bigg\{ ~~ 
D(k^2)
 \Bigg(
     A(k^2) 
      \Bigg[ ~  3 \, \tau_5 \, k^2 ~ + ~   \frac{3 }{2} \, \tau_7 \, k^4
      \Bigg]
 ~ + ~ B(k^2)  
           \Bigg[
              3 \, \lambda_1
              ~ + ~ 3 \,  \tau_3 \,  k^2
              ~ - ~ 3 \, \tau_6 \, k^2 
            \Bigg] ~~
\Bigg)
\nonumber \\
& & \qquad\qquad
+ ~ \frac{\xi}{k^2} 
    \Bigg(
       A(k^2)  \Bigg[ ~ - \,   \lambda_3 \,  k^2  \Bigg]
       ~ + ~ B( k^2) \Bigg[  \lambda_1 +  \lambda_2 \,  k^2   \Bigg]
    \Bigg)
    ~\Bigg\} 
      \label{DSE-Fermion-scalar-zero}
\end{eqnarray}
and $B(0)$ is determined by the longitudinal form factors and the $\tau_3$, $\tau_5$, $\tau_6$ and $\tau_7$. In what concerns chiral symmetry
breaking either in QED or in QCD any realistic vertex successful model should contain at least one of these transverse form factors.
The analysis of the vector component of the DSE is slightly more complicated as Eq. (\ref{DSE-Fermion-scalar}) has to be divided by $p^2$ and only then
the limit $p \rightarrow 0$ should be taken. As long as $B \ne 0$ and the photon propagator does not diverge as in the perturbative result, it seems that
there are no IR divergences for $B(0)$.

\subsection{The photon gap equation \label{Sec:PropEqPhoton}}

Let us proceed with the analysis of the DSE for the photon (\ref{Eq:DSE-Photon}). Using the tensor basis described previously, then
the computation of the trace of the fermion bubble results in
\begin{eqnarray}
& &
  \frac{1}{D(k^2)}  = 
   k^2 
  - i \, \frac{g^2}{3} \,  \int \frac{d^4 p}{(2 \, \pi)^4} ~  \frac{1}{A^2(p^2) \, p^2 - B^2(p^2)} ~ \frac{1}{A^2((p-k)^2) \, (p-k)^2 - B^2((p-k)^2)} \Bigg\{
  \nonumber \\
  & & 
A(p^2) A((p-k)^2) 
\Bigg[
 8 \, \lambda_1 \bigg( (pk) -  p^2\bigg)
 ~ + ~ 4 \, \lambda_2 \bigg(  3 \, k^2 \, p^2 + 4 \, p^4 -  (k^2 + 8 \, p^2) (pk) + 2 \, (pk)^2 \bigg)
 \nonumber \\
 & & \hspace{3cm}  
 + ~ 4 \, \tau_2 \bigg(  k^4  \, p^2 + 2  \, k^2  \, p^4 - 3 \, k^2 \, p^2 \, (pk) -  ( k^2 + 2 \, p^2) \, (pk)^2 + 2 (pk)^3  \bigg)
   \nonumber \\
   & & \hspace{3cm}                    
   + ~ 4 \, \tau_3 \bigg( -   \, k^2  \, p^2 + 3  \, k^2  \,  (pk) - 2  \, (pk)^2\bigg)
   \nonumber \\
   & & \hspace{3cm}                    
   + ~ 12 \, \tau_6 \bigg(  - \,  k^2 \, p^2 + \, ( k^2 + 2 \, p^2) (pk) - 2 \, (pk)^2 \bigg)
   ~ + ~8 \,  \tau_8 \bigg( k^2 \, p^2 - (pk)^2  \bigg)
\Bigg]
\nonumber \\
& &
+  A(p^2) B((p-k)^2) 
\Bigg[ 
   4 \,  \lambda_3 \bigg((pk)-2 p^2\bigg) 
   ~ + ~ 4 \, \tau_1 \bigg( k^2 \, p^2 -  (pk)^2)\bigg)
  ~ + ~ 4 \, \tau_4 \, (pk) \, \bigg( k^2 \, p^2 - (pk)^2 \bigg)
   \nonumber \\
   & & \hspace{3cm}                    
~ + ~12 \, \tau_5 \, (pk)
   ~ + ~  \tau_7 \bigg(  - \, 8 \, k^2 \, p^2 +  ( 6 \,  k^2 + 24 \, p^2) \, (pk) - 16 \, (pk)^2 \bigg)
\Bigg]
\nonumber \\
& &
+B(p^2) A((p-k)^2) 
\Bigg[
   4 \, \lambda_3 \bigg(  - 2 \, p^2 -  k^2 + 3 \, (pk)\bigg)
   ~ + ~ 4 \, \tau_1 \bigg( k^2 \, p^2 -  (pk)^2 \bigg)
 \nonumber \\
 & & \hspace{3cm}               
 + ~ 4 \, \tau_4 \bigg( k^4 \,  p^2  -  k^2 \,  p^2 \,  (pk) - k^2 \, (pk)^2  +  \, (pk)^3\bigg)
 ~ + ~ 12 \, \tau_5 \bigg(k^2 - (pk)\bigg)
 \nonumber \\
 & & \hspace{3cm}                 
 + ~ \tau_7 \bigg(    6 \, k^4 + 16 \, k^2 \, p^2 - 6 \, ( 5 \, k^2 + 4 \, p^2 ) \, (pk) + 40 \, (pk)^2 - 8 \, (pk)^3 \bigg)
\Bigg]
\nonumber \\
& &
+B(p^2) B((p-k)^2) 
\Bigg[
16 \, \lambda_1
~ +~ 4 \, \lambda_2 \bigg( k^2 + 4 \, p^2 - 4 \, (pk) \bigg)
~ + ~ 8 \, \tau_2 \bigg(  k^2 \, p^2 - (pk)^2\bigg)
  \nonumber \\
 & & \hspace{3cm}  
  + ~ 12 \, \tau_3 \, k^2
 ~ + ~ 12 \, \tau_6 \bigg( k^2 - 2 \, (pk) \bigg)
\Bigg] ~
\Bigg\} \ ,
     \label{Eq:PhotonEqBasis}
\end{eqnarray}
where the form factors are now
\begin{equation}
\lambda_i = \lambda_i \big( (p-k)^2, \, p^2, \, k^2\big)
\qquad\mbox{ and }\qquad 
\tau_i = \tau_i \big( (p-k)^2, \, p^2, \, k^2\big) \ .
\end{equation}
As for the fermion gap equation, the photon DSE gets contributions from all the transverse form factors. It is only for chiral theories where $B = 0$,
that a subset of the vertex form factors contributes to this integral equation.

The photon propagator DSE takes a relative simple form when the photon momentum vanish. Indeed, for $k = 0$ the transverse operators describing
the photon-fermion vertex do not contribute to $D(0)$ and the exact bare equation becomes
\begin{eqnarray}
& &
  \frac{1}{D(0)}   ~ = ~
 - i \, \frac{8 }{3} \, g^2 \,  \int \frac{d^4 p}{(2 \, \pi)^4} ~  \frac{1}{\bigg( A^2(p^2) \, p^2 - B^2(p^2)\bigg)^2} 
  \nonumber \\
  & & 
  \Bigg\{
    A(p^2) \bigg( - \, A^2(p^2) \, p^2 + 2 \, B^2(p^2) \bigg)
   ~ + ~  A^\prime(p^2) \, p^2 \, \bigg( A^2(p^2) \, p^2 + B^2(p^2) \bigg)
   ~ - ~ 2 \, B^\prime(p^2) \, p^2 \, A(p^2) \, B(p^2) 
  \Bigg\} \ .
     \label{Eq:PhotonEqBasis-zeromom}
\end{eqnarray}
If the photon is a massless particle the integral appearing in this last equation has to vanish, which imposes non-trivial constraints
to the fermion propagator functions $A$ and $B$.

\subsection{The photon-fermion vertex equation \label{SecVertex}}

Let us   now return to the  Dyson-Schwinger equation for the photon-fermion vertex as given in Eq. (\ref{Eq:DSE-vertex}).
Using, once more, the tensor basis defined in Eqs (\ref{TensorBasis-Long}) and (\ref{TensorBasis-Ortho}),  then the l.h.s. of this integral equation
can be written (highlighting the tensor decomposition of the Dirac forms) as
\begin{eqnarray}
& & 
{\Gamma}^\mu (p, \, -p -k; \, k)   ~ = ~ 
     {p}^{\mu } ~  \Big(\, {k}^2 \,  {\tau_1} \,  - \, 2 \, {\lambda_3} \, \Big) - {k}^{\mu }  ~ \Big(\, (pk) \,  {\tau_1} \,  + \, {\lambda_3} \Big)
     \hspace{1.5cm}\text{(Scalar Component)}
     \nonumber \\
     & & 
     \qquad\qquad
     + \, 
     {\gamma }^{\mu } \Bigg[ \, {\lambda_1} \, + \, k^2 \, {\tau_3}+\,  \Big( k^2  \, + \, \, 2 \,  ( {p} {k}) \Big) \, {\tau_6} \, \Bigg]
      \hspace{3.6cm}\text{(Vector Components)}
     \nonumber \\
     & & \qquad\qquad\qquad
     + \, \slashed{k}  \, \Bigg[ \, {k}^{\mu } \Big( \, {\lambda_2} \, - \,  ({p} {k} ) \,  {\tau_2} \, - \, {\tau_3} \, -\, {\tau_6}\Big) 
                                             \, + \, {p}^{\mu } \Big(2 \, {\lambda_2} \, + \, k^2 \, \, {\tau_2}- \, 2 \, {\tau_6} \Big) \, \Bigg]
     \nonumber \\
     & & \qquad\qquad\qquad
     + \, 2 \, \slashed{p} \, \Bigg[ \, {k}^{\mu } \Big(\, {\lambda_2} \, -  \, ({k} {p}) \, {\tau_2}\Big) \, + \, {p}^{\mu } \Big( 2 \, {\lambda_2} \, + \, k^2 \, {\tau_2} \Big)\,  \Bigg]
     \nonumber \\
     & & \qquad\qquad
     + \, i ~~ {\tau_8}  ~~ {\gamma }_{\sigma } ~ {\gamma }_5 ~~ ~{\epsilon }^{\sigma  \mu  \alpha\beta} ~~  {p}_\alpha \, {k}_\beta
           \hspace{5.3cm}\text{(Axial Component)}
     \nonumber \\
     & & \qquad\qquad
     + \,  
     \frac{1}{2} \, \sigma^{\mu\alpha} \, k_\alpha \,  \Bigg[ 2 \, \tau_5 \, + \, \tau_7 \Big( k^2 + 2 \, (kp) \Big) \Bigg]
     \hspace{4.cm}\text{(Tensor Components)}
     \nonumber \\
     & & \qquad\qquad\qquad
     + \,  
          \sigma^{\mu\alpha} \, p_\alpha \,  \Bigg[ \,  \tau_7 \Big( k^2 + 2 \, (kp) \Big) \, \Bigg]
     \nonumber \\
     & & \qquad\qquad\qquad
     \, + \,  \sigma^{\alpha\beta} \, p_\alpha \, k_\beta \Bigg[ \, k^\mu \, \Big( \tau_7 -  (pk) \, \tau_4  \Big) +  p^\mu \Big(  k^2 \, \tau_4  +  2 \, \tau_7  \Big) \Bigg]
     \label{FullVertex-Dirac}
\end{eqnarray}
where the form factors are
\begin{equation}
   \lambda_i = \lambda_i \Big((p+k)^2, \, p^2, \, k^2\Big)
   \qquad\mbox{ and }\qquad
   \tau_i = \tau_i \Big((p+k)^2, \, p^2, \, k^2\Big) \ .
\end{equation}   
Note that, besides the longitudinal contribution associated with the various $\lambda_i$, that are fixed by the Ward-Takahashi identity, the Dirac
scalar term requires only $\tau_1$,  the Dirac vector term calls for $\tau_2$, $\tau_3$ and $\tau_6$,
the Dirac tensor involves $\tau_4$, $\tau_5$, $\tau_7$ and the axial vector terms is described by a single form factor $\tau_8$.  
A cross-check of the decomposition given in Eq. (\ref{FullVertex-Dirac}) is to compute $k_\mu \Gamma^\mu$ using (\ref{FullVertex-Dirac}) and, indeed,
a direct inspection shows that only the longitudinal form factors appear in this scalar form of the equation as expected.

It is common practise to take models for the vertices when considering either QED or QCD that simplify the outcome of Eq. (\ref{FullVertex-Dirac}). 
Note that this equation is the decomposition in terms of Dirac bilinears of the vertex and, module color structures and the $\lambda_4$ contributioin, 
is valid for the Abelain and for the non-Abelian theory.
A subset of the available models can be found in  
\cite{Roberts:1994dr,Maris:1999nt,Kizilersu:2009kg,Kizilersu:2014ela,Lessa:2022wqc} and references therein. 
For example, for the Curtis-Pennington vertex \cite{Curtis:1990zs,Curtis:1992jg} the vertex equation (\ref{FullVertex-Dirac}) reduces to
\begin{eqnarray}
{\Gamma}^\mu (p, \, -p -k; \, k)   & =  & 
       \Big(\,  - \, 2 \, {p}^{\mu } ~  - ~  {k}^{\mu } \, \Big)  ~ {\lambda_3} 
     ~ + ~ 
     {\gamma }^{\mu } \Bigg( \, {\lambda_1} \, + \,  \Big( k^2  \, + \,  2 \,  ({k} {p}) \, \Big) \, {\tau_6}  \, \Bigg) 
     \nonumber \\
     & & \qquad
     + \, \slashed{k}  \, \Bigg( \, {k}^{\mu }  \, + \, 2 \, {p}^{\mu }  \, \Bigg) 
                                             \Big( \, {\lambda_2}  \, -\, {\tau_6}\Big) 
    ~  + ~ 2 \, \slashed{p} \, \Bigg( \, {k}^{\mu }  \, + \, 2 \, {p}^{\mu }  \,  \Bigg)
     \, {\lambda_2} \ .
     \label{FullVertex-Dirac-CP}
\end{eqnarray}
This vertex model in particular considers only a single transverse form factor $\tau_6$ that also contributes to $B(0)$ and, therefore, can be relevant also
for the mechanics of chiral symmetry breaking.

The contraction of the vertex equation with the incoming fermion momentum $p^\mu$ results in another Lorentz scalar that can
help in the computation of the form factors. Indeed, given that 
\begin{eqnarray}
& & 
p_\mu \,{\Gamma}^\mu (p, \, -p -k; \, k)   ~ = ~ 
     \tau_1 ~ \Big( \, p^2 \, k^2 - (pk)^2 \,\Big) \, - \, \lambda_3 ~ \Big( \, 2 \, p^2 + (pk) \, \Big) 
     \nonumber \\
     & & 
     \qquad\qquad
     + \, 
     \slashed{k} \Bigg[ \, 
              \lambda_2 \bigg( (kp) + 2 \, p^2 \bigg)
              ~ + ~ \tau_2 \bigg( k^2p^2 - (kp)^2 \bigg)
              ~ - ~ \tau_3 \, (kp) 
              ~ - ~ \tau_6 \bigg( (kp) + 2 \, p^2 \bigg)
      \, \Bigg]
     \nonumber \\
     & & \qquad\qquad\qquad
     + \,  \slashed{p} \, \Bigg[ \, 
          \lambda_1
          ~ + ~ 2 \, \lambda_2 \bigg( (kp) + 2 \, p^2 \bigg)
          ~ + ~ 2 \, \tau_2 \bigg( k^2 p^2 - (kp)^2 \bigg)
          ~ + ~ \tau_3 \, k^2 
          ~ + ~ \tau_6 \bigg( k^2 + 2 \, (kp) \bigg) 
     \,  \Bigg]
     \nonumber \\
     & & \qquad\qquad
     + \,  
     \sigma^{\alpha\beta} \, p_\alpha \, k_\beta \Bigg[ \, 
      \tau_4 \bigg( p^2 k^2 - (kp)^2 \bigg)
      ~ + ~ \tau_5 
      ~ + ~ \tau_7 \bigg( \frac{1}{2} k^2 + 2 \, (kp) +  2 \, p^2 \bigg)       
     \, \Bigg]
     \label{FullVertex-Dirac-Contraction}
\end{eqnarray}
and taking into account that the longitudinal form factors are determined by solving the Ward-Takahashi identity for the vertex, then  $\tau_1$
can be determined solving the equation
\begin{equation}
  \tau_1  \Big((p+k)^2, \, p^2, \, k^2\Big)=   ~ \frac{1}{ p^2 \, k^2 - (pk)^2 } \Bigg( 
\frac{1}{4} \, \text{Tr} \bigg( p_\mu \,{\Gamma}^\mu (p, \, -p -k; \, k)   \bigg) +  \lambda_3 ~ \Big( \, 2 \, p^2 + (pk) \, \Big) 
\Bigg) \ .
\label{tau1-fromvertex}
\end{equation}   
This is an exact relation, that also applies to QCD, and is an example of what we call exact kinematical equations that come from
exploring the tensor properties of $\Gamma^\mu$.
For a tree level type of vertex where $\Gamma^\mu = \gamma^\mu$ and, therefore, $\lambda_3 = 0$, it follows that $\tau_1 = 0$ as expected.
Another example of an exact kinematical equation for the photon-fermion vertex is related to the computation of $\tau_8$ that that can be calculated by solving
\begin{equation}
~~ {\tau_8}  \Big((p+k)^2, \, p^2, \, k^2\Big) ~~{\epsilon }^{\zeta  \mu  \alpha\beta} ~~  {p}_\alpha \, {k}_\beta ~ = ~
i ~~ \text{Tr} \Bigg( {\Gamma}^\mu (p, \, -p -k; \, k)   \, \gamma^\zeta \, \gamma_5 \Bigg)  \ .
\end{equation}
This form factor also vanishes for a tree level type of vertex.

The vertex equation given in (\ref{FullVertex-Dirac}) hold for any kinematics. However, by choosing a particular set of $p$ and $k$ the equation can
become simpler. A particular kinematic configuration that simplifies the vertex DSE occurs when the incoming fermion momentum $p$ vanish and,
in this case,
\begin{eqnarray}
{\Gamma}^\mu (0, \, -k; \, k)   & = &
     - {k}^{\mu }  ~  {\lambda_3} 
    ~  + ~ 
     {\gamma }^{\mu } \Bigg[ \, {\lambda_1} \, + \, k^2 \, \Big( \tau_3   \, + \,  {\tau_6} \Big) \, \Bigg]
     ~ + ~ \slashed{k}  ~ {k}^{\mu } \Bigg[ \, {\lambda_2}  \, - \, {\tau_3} \, -\, {\tau_6}\Bigg]
     \nonumber \\
     & & \qquad\qquad
     + \,  
     \frac{1}{2} \, \sigma^{\mu\alpha} \, k_\alpha \,  \Bigg[ 2 \, \tau_5 \, + \, k^2  \, \tau_7  \Bigg] \ .
     \label{FullVertex-Dirac-pzero}
\end{eqnarray}
For $ p = 0$ it is enough to know the combinations of transverse form factors $\tau_3 + \tau_6$ and $2 \, \tau_5 \, + \, k^2  \, \tau_7$.
Another kinematical configuration where the equation simplifies considerably is when the photon momentum vanish (soft photon limit).
In this case only the longitudinal form factors $\lambda_i$ contribute to the vertex DSE. Moreover, when the incoming fermion momentum and
the photon momentum are orthogonal, i.e. when $(pk) = 0$, further simplifications take place and
\begin{eqnarray}
& & 
{\Gamma}^\mu (p, \, -p -k; \, k)   ~ = ~ 
     {p}^{\mu } ~  \Big(\, {k}^2 \,  {\tau_1} \,  - \, 2 \, {\lambda_3} \, \Big) - {k}^{\mu }  ~  {\lambda_3} 
     \nonumber \\
     & & 
     \qquad\qquad
     + \, 
     {\gamma }^{\mu } \Bigg[ \, {\lambda_1} \, + \, k^2 \, \Big( {\tau_3} \, + \,   {\tau_6} \Big) \, \Bigg]
     ~ + ~ \slashed{k}  \, \Bigg[ \, {k}^{\mu } \Big( \, {\lambda_2} \, - \, {\tau_3} \, -\, {\tau_6}\Big) 
                                             \, + \, {p}^{\mu } \Big(2 \, {\lambda_2} \, + \, k^2 \, \, {\tau_2}- \, 2 \, {\tau_6} \Big) \, \Bigg]
     \nonumber \\
     & & \qquad\qquad\qquad
     + \, 2 \, \slashed{p} \, \Bigg[   {p}^{\mu } \Big( 2 \, {\lambda_2} \, + \, k^2 \, {\tau_2} \Big)\, \, - \,  {k}^{\mu } \, {\lambda_2}  \Bigg]
     \nonumber \\
     & & \qquad\qquad
     + \, i ~~ {\tau_8}  ~~ {\gamma }_{\sigma } ~ {\gamma }_5 ~~ ~{\epsilon }^{\sigma  \mu  \alpha\beta} ~~  {p}_\alpha \, {k}_\beta
     \nonumber \\
     & & \qquad\qquad
     + \,  
     \frac{1}{2} \, \sigma^{\mu\alpha} \, k_\alpha \,  \Bigg[ 2 \, \tau_5 \, + \, \tau_7 \, k^2  \Bigg]
      ~ + ~  
          \sigma^{\mu\alpha} \, p_\alpha \,  \Bigg[ \,  \tau_7 \, k^2  \, \Bigg]
     ~  + ~ \sigma^{\alpha\beta} \, p_\alpha \, k_\beta \Bigg[ \, k^\mu \, \tau_7  +  p^\mu \Big(  k^2 \, \tau_4  +  2 \, \tau_7  \Big) \Bigg] \ .
     \label{FullVertex-Dirac-orthogonal}
\end{eqnarray}

The decomposition of the vertex given in Eq. (\ref{FullVertex-Dirac}) or any of its simplifications, see Eqs (\ref{FullVertex-Dirac-pzero}) and
(\ref{FullVertex-Dirac-orthogonal}), rely only on the tensor decomposition of $\Gamma^\mu$ and, therefore, are valid both for
the photon-fermioin vertex and also to the quark-gluon vertex, module its color structure and the contribution associated with $\lambda_4$. 
The same rationale applies to the decomposition of the fermion gap equation in terms of the vertex form factors, 
see Eqs (\ref{DSE-Fermion-scalar}) and (\ref{DSE-Fermion-vector}). Indeed, the QCD fermion gap equation, after color integration and the contribution
associated with $\lambda_4$ that changes only the scalar equation, is the 
same up to a proper replacement of the coupling constant and the renormalization constants. The main difference between QCD and QED
is the bosonic Dyson-Schwinger equations that in QCD requires triple and quartic gluonic vertices and also ghost contributions.

The computation of the  transverse form factors from the vertex DSE (\ref{FullVertex-Dirac}) requires the dynamical information that is contained in the
r.h.s. side of the vertex equation. Handling exact expressions on the r.h.s. of this equation is complex and it is helpful to introduce simplifications and,
therefore, model the contributions appearing there. We will be postponed to Sec.\ref{SecTransverseFF} the derivation of exact kinematical
expressions for each of the transverse form factors $\tau_i$.

\section{A  model to feed the r.h.s. of the vertex Dyson-Schwinger equation \label{Sec:rhsVertex}}

The computation of a solution of the vertex Dyson-Schwinger equation (\ref{Eq:DSE-vertex}) requires, besides the fermion functions $A$ and
$B$, the two-photon-two-fermion vertex $\Gamma^{\mu\nu}$. 
The determination of the two-photon-two-fermion vertex is challenging and even if such vertex is known, 
computing a solution for the vertex from its DSE is also arduous and demands a complex numerical procedure.
Even disregarding the two-photon-two-fermion contribution to the equation, inserting in the r.h.s side of the equation the full vertex
results in a lengthy expression that is difficult to handle and analyse. In this approximation the vertex is a non-linear integral equation
that is not simple to solve.
However, by simplifying the terms appearing in the integrals of the r.h.s. of Eq. (\ref{Eq:DSE-vertex}), i.e. using on 
its r.h.s simplified $\Gamma^\mu$ and $\Gamma^{\mu\nu}$ vertices, one can get some insight on $\Gamma^\mu$. 

The usual approach to  QED is based on perturbation theory and, therefore, the perturbative approach will always be in back of our mind.
Moreover, calling for a simplified model to feed the r.h.s. of the vertex equation it is possible to decompose the r.h.s. of Eq. (\ref{Eq:DSE-vertex}) into
the Dirac bilinear algebra basis and, combined with Eq. (\ref{FullVertex-Dirac}), build approximate expressions for the various transverse
form factors that go beyond their perturbative estimation. 

Let us discuss the building of a vertex model to feed the vertex DSE.
The photon propagator depends on the gauge and introduces, in the vertex equation, a dependence on $\xi$. By working in the Landau gauge, i.e. 
by setting $\xi = 0$, the number of terms in the r.h.s. of the equation will be reduced. Moreover, the longitudinal vertex form factors are
fixed by the vertex Ward-Takahashi identity and, therefore, a ``natural'' choice is to consider only the pure longitudinal vertex in the r.h.s.
The solutions of the Ward-Takahashi identity given in Eqs (\ref{EQ:L1}) to (\ref{EQ:L4}) allow to set the relative importance
of $\lambda_1$, $\lambda_2$ and $\lambda_3$. For example, if $A$ and $B$ are independent of the momentum, or almost independent
of momentum as occurs in the perturbative solution of QED, then the only non-vanishing longitudinal form factor contributing in the r.h.s. of the
equation is $\lambda_1$. A possible choice is, in a first approximation, to ignore the contributions proportional to $\lambda_2$ and $\lambda_3$. 
Note, however, that the contribution of $\lambda_3$  can be sizeable if $B$ has a strong dependence on the $p^2$ as happens in QCD 
due to dynamical chiral symmetry breaking \cite{Oliveira:2018ukh,Lessa:2022wqc}.
Although chiral symmetry breaking can occur in QED for sufficiently large coupling, we will not consider this case here.

The perturbative solution of the vertex DSE is constructed setting in the r.h.s. $\Gamma^\mu = \gamma^\mu$, then, after performing the
momentum integration, get $\Gamma^\mu$ and used it to feed back the r.h.s., repeating the calculation until convergence is achieved. 
The full perturbative solution implies also an iteration process for the propagator equations. Then, by taking into account in the r.h.s. of the 
vertex equation (\ref{Eq:DSE-vertex}) only the contribution associated with $\lambda_1$ but incorporating its full momentum dependence,
the analysis of $\Gamma^\mu$ goes already beyond its perturbative solution. Furthermore, by adding a contribution coming from the 
longitudinal component of the two-photon-two-fermion vertex $\Gamma^{\mu\nu}$, that vanishes in first order in perturbation theory, once more 
our approach goes beyond the perturbative analysis of the vertex equation.

To resume, whenever it will required to ``solve'' the vertex equation only the contributions associated with
the tree level tensorial structure for the photon-fermion vertex will be considered,
together with the $\Gamma^{\mu\nu}$ that solves its WTI as computed in \cite{Oliveira:2022bar}.
Furthermore, the vertex equation will be worked out in the Landau gauge. To simplify the analysis of the vertex equation, we will
treat separately the contribution that includes $\Gamma^{\mu\nu}$ and the remaining term, that will be named pure vertex contribution.
Let us mention that by considering only the longitudinal vertex and, in particular, only $\lambda_1$ in the r.h.s. of the vertex DSE gauge
symmetry is not preserved exactly but, at most, is only satisfied approximately.

\subsection{Contributions from the Pure Vertex}

In order to simplify the writing of the pure vertex contribution, before performing its Dirac decomposition, let us introduce some new notation 
and define
\begin{eqnarray}
& & 
\MyInt  =  ~
 - ~ i \, g^2 \, \int \frac{d^4 q}{(2 \, \pi)^4} ~ D(q^2) ~ \frac{1}{ A^2 \Big( (p-q)^2 \Big) ~ (p-q)^2 ~ - ~  B^2 \Big( (p-q)^2 \Big) } \nonumber \\
 & & \qquad\qquad\qquad\qquad\qquad\qquad\qquad\qquad
 \frac{1}{A^2 \Big( (p+k-q)^2 \Big) ~ (p+k-q)^2 ~ - ~  B^2 \Big( (p+k-q)^2 \Big) } \ .
\end{eqnarray}
Then, within the setup discussed above, the pure vertex contribution of the r.h.s of the vertex equation reads
\begin{eqnarray}
& & 
\MyInt  ~~ \lambda_1 \big( (p+k-q)^2, \, (p-q)^2, \, k^2\big) ~~ \lambda_1\big( (p+k)^2, \, (p+k-q)^2, \, q^2 \big)
\nonumber \\
& & \hspace{0.6cm}
\Bigg\{~ 
  3 \, p^\mu \Big[ A\big( (p-q)^2\big) ~   B\big( (p+k-q)^2 \big) ~ + ~ B\big( (p-q)^2\big)  ~  A\big( (p+k-q)^2 \big) \Big]  
      \nonumber \\  
      & & \hspace{3cm} 
       ~ - ~ 3 \, q^\mu \Big[ A\big( (p-q)^2\big) ~   B\big( (p+k-q)^2 \big) ~ + ~ B\big( (p-q)^2\big)  ~  A\big( (p+k-q)^2 \big) \Big] 
      \nonumber \\  
      & & \hspace{3cm} 
      ~ + ~ 3\, k^\mu  \Big[  B\big( (p-q)^2\big)  ~  A\big( (p+k-q)^2 \big) \Big] 
\nonumber \\
& & \hspace{0.9cm}
  + ~ \gamma^\mu \Bigg[ A\big( (p-q)^2\big) ~ A\big( (p+k-q)^2 \big) \Bigg( (kp) - (kq) - 2 \,  (pq) + p^2 + q^2 \Bigg)
                                    ~ - ~ B\big( (p-q)^2\big) ~  B\big( (p+k-q)^2 \big)
                              \Bigg]
\nonumber \\
& & \hspace{3cm}
 ~ + ~ A\big( (p-q)^2\big) ~ A\big( (p+k-q)^2 \big)  
          \Bigg[ ~ 
            ~ k^{\mu } \slashed{q} \Bigg( 3 ~ - ~ 2 \,  \frac{(pq) }{{q}^2} \Bigg)  ~ - ~ k^{\mu } \slashed{p} ~~ - ~ p^{\mu } \slashed{k} 
           ~ - ~ 2 \, p^{\mu } \slashed{p} 
           \nonumber \\
           & & \hspace{8.5cm}           
           ~ + ~ 2 \, p^{\mu } \slashed{q} \Bigg( 3 ~  -  ~  \frac{ (kq) + 2 \, (pq) }{{q}^2}   \Bigg)
            ~ + ~ q^{\mu } \slashed{k} ~ + ~ 2 \, q^{\mu } \slashed{p}  ~
\Bigg]
\nonumber \\
& & \hspace{3.5cm} 
   ~ + ~ 2 \, q^{\mu } \slashed{q} \Bigg[
             A\big( (p-q)^2\big) ~ A\big( (p+k-q)^2 \big)  \Bigg(  - \, 2 ~ + ~   \frac{ (kp) + p^2}{{q}^2}  \Bigg)
                                 \nonumber \\
                                 & & \hspace{7cm} 
             ~ - ~ \frac{ B\big( (p-q)^2\big) ~  B\big( (p+k-q)^2 \big)}{q^2}  \Bigg]
\nonumber \\
& & \hspace{0.9cm}
 ~ - ~  i ~ A\big( (p-q)^2\big) ~  A\big( (p+k-q)^2 \big) 
    \nonumber \\
    & & \hspace{2cm}
     \Bigg[  \Bigg( 2 \,   \frac{(pq)}{{q}^2}  + 1 \Bigg) ~  \epsilon^{\mu \alpha \beta \eta} ~ k_\alpha \, q_\beta 
    ~ - ~ 2 \,  
    \frac{ (kq)}{{q}^2}  ~ \epsilon ^{\mu \alpha \beta \eta} ~ p_\alpha q_\beta
    ~ - ~ 3  ~  \epsilon^{\mu \alpha \beta \eta} ~ k_\alpha p_\beta 
    \Bigg] \, \gamma_\eta \, \gamma_5
\nonumber \\
& & \hspace{0.9cm}
 ~ - ~ \sigma^{\mu\alpha} k_\alpha \Bigg[ B\big( (p-q)^2\big) ~  A\big( (p+k-q)^2 \big) \Bigg]
\nonumber \\
& & \hspace{2cm}
 ~ + ~
     \sigma^{\mu\alpha} p_\alpha \Bigg[ A\big( (p-q)^2\big) ~  B\big( (p+k-q)^2 \big)  ~ -  ~ B\big( (p-q)^2\big) ~  A\big( (p+k-q)^2 \big) \Bigg]
\nonumber \\
& & \hspace{2.5cm}
 ~ + ~\sigma^{\mu\alpha} q_\alpha \Bigg[
             A\big( (p-q)^2\big) ~  B\big( (p+k-q)^2 \big) \Bigg( 1 - 2 \, \frac{(pq)}{q^2} \Bigg) 
             \nonumber \\
             & & \hspace{6cm}
             ~ + ~ B\big( (p-q)^2\big) ~  A\big( (p+k-q)^2 \big)\Bigg( -1 + 2 \, \frac{(pq) + (kq)}{q^2} \Bigg)
 \Bigg]
\nonumber \\
& & \hspace{3cm}
  ~ + ~  2 \, \frac{q^\mu}{q^2} \, \sigma^{\alpha\beta}p_\alpha q_\beta 
       \Bigg[ A\big( (p-q)^2\big) ~  B\big( (p+k-q)^2 \big) ~ - ~ B\big( (p-q)^2\big) ~  A\big( (p+k-q)^2 \big)
       \Bigg]
\nonumber \\
& & \hspace{3.5cm}
 ~ - ~ 2 \, \frac{q^\mu}{q^2} \, \sigma^{\alpha\beta} k_\alpha q_\beta \Bigg[  B\big( (p-q)^2\big) ~  A\big( (p+k-q)^2 \big)
   ~ \Bigg] ~
\Bigg\} \ ,
\label{EqRhSVertex}
\end{eqnarray}
where the outcome was organized according to its bilinear Dirac algebra tensor properties and
\begin{eqnarray}
\lambda_1 \big( (p+k-q)^2, \, (p-q)^2, \, k^2\big) & = & 
     \frac{1}{2} \Bigg( A\Big( (p+k-q)^2 \Big) + A\Big((p-q)^2\Big) \Bigg)
\\
\lambda_1\big( (p+k)^2, \, (p+k-q)^2, \, q^2 \big) & = & 
     \frac{1}{2} \Bigg( A\Big( (p+k-q)^2 \Big) + A\Big((p+k)^2\Big) \Bigg)
\end{eqnarray}
are the solutions of the vertex WTI. Despite using a simplified vertex in the r.h.s of the equation, its structure generate contributions to all
the Dirac bilinear forms.
Note also, that for a chiral theory where $B = 0$, there are huge simplifications and the only non-vanishing
terms are those for the Dirac vector and axial vector bilinear forms.

\subsection{The Longitudinal Two-Photon-Two-Fermion Contribution}

Let us now discuss the case of the contribution that is associated with the two-photon-two-fermion vertex. In its evaluation we take a similar approach  as
before and consider only the longitudinal part of $\Gamma^{\mu\nu}$, i.e. the solution of the Ward-Takahashi for the two-photon-two-fermion 
vertex. The solution of the WTI was determined in \cite{Oliveira:2022bar} and reads
\begin{eqnarray}
   \Gamma^{\mu\nu}_L (p, \, -p-k-q; \, k, \, q)  & = &  \frac{k^\mu}{k^2} \bigg( \Gamma^\nu ( p, \, - p - q; \, q )  ~ -   ~\Gamma^{\nu} ( p+k, \, -p-k-q; \, q) \bigg)  \nonumber \\
   & & \qquad + ~  \frac{q^\nu}{q^2} \bigg( \Gamma^\mu ( p, \, - p - k; \, k )  ~ -   ~\Gamma^{\mu} ( p+q, \, -p-k-q; \, k) \bigg) \nonumber \\
   & & \qquad\qquad + ~\widetilde{A}(k,q) \left(  g^{\mu\nu} \, - \,  \frac{k^\mu k^\nu}{k^2} \, - \, \frac{q^\mu q^\nu}{q^2}  \right)  
   \label{Sol:WTITwoPhotonTwoFermion1}
\end{eqnarray}         
where the symmetric form factor in the last line is given by
\begin{eqnarray}
  \widetilde{A}(q,k) 
         & = & \frac{1}{ (k \cdot q)} \Bigg\{
                \bigg[ A\left( (p+k)^2\right) + A\left( (p+q)^2\right) - A\left( (p+k+q)^2\right) - A\left( p^2\right) \bigg] ~\slashed{p} \nonumber \\
                & & \qquad\qquad + ~  \bigg[ A\left( (p+k)^2\right)  - A\left( (p+k+q)^2\right) \bigg] ~\slashed{k} \nonumber \\
                & & \qquad\qquad + ~  \bigg[ A\left( (p+q)^2\right)  - A\left( (p+k+q)^2\right) \bigg] ~\slashed{q} \nonumber \\                
                & & \qquad\qquad +  ~ B\left( (p+k+q)^2\right) + B\left( p^2\right) -  B\left( (p+k)^2\right) - B\left( (p+q)^2\right) \Bigg\} 
           \label{Sol:WTITwoPhotonTwoFermion3}
 \end{eqnarray}
 By definition, the function $\widetilde{A}$ is given by differences and  sums of differences of fermion propagator functions
 $A$ and $B$ at different momenta and inserting the lowest order, in the coupling constant, solution for $A$ and $B$ obtained within perturbation 
 theory this term  vanish. 
 If, as for the previous term in the vertex DSE, only the $\lambda_1$ term is taken into account whenever $\Gamma^\mu$ appears, then 
 \begin{eqnarray}
   \Gamma^{\mu\nu}_L (p, \, -p-k-q; \, k, \, q)  & = &  
   \frac{k^\mu \, \gamma^\nu }{k^2} \bigg(  \lambda_1 \big( (p + q)^2, \, p^2, \, q^2 \big)  ~ - ~ \lambda_1 \big( (p+k+q)^2, \, (p+k)^2, \, q^2 \big) \bigg)  
   \nonumber \\
   & & \qquad 
   + ~  \frac{q^\nu \, \gamma^\mu }{q^2} \bigg( \lambda_1 \big( (p+k)^2, \, p^2, \, k^2 \big)  ~ - ~ \lambda_1 \big( (p+k+q)^2, \, (p+q)^2 , \, k^2 \big) \bigg) 
   \nonumber \\
   & & \qquad\qquad + ~\widetilde{A}(k,q) \left(  g^{\mu\nu} \, - \,  \frac{k^\mu k^\nu}{k^2} \, - \, \frac{q^\mu q^\nu}{q^2}  \right)  
   \nonumber \\
   & = &  
   \frac{1}{2} \, \frac{k^\mu \, \gamma^\nu }{k^2} \bigg(  A\big( (p + q)^2\big) ~  + ~ A \big( p^2 \big)   ~ - ~ A \big( (p+k+q)^2 \big) ~ - ~  A \big( (p+k)^2 \big) \bigg)  
   \nonumber \\
   & & \qquad 
   + ~   \frac{1}{2} \,\frac{q^\nu \, \gamma^\mu }{q^2} \bigg(  A \big( (p+k)^2 \big)  ~ + ~ A \big( p^2 \big)  ~ - ~ A \big( (p+k+q)^2 \big) ~ -  ~ A \big( (p+q)^2  \big) \bigg) 
   \nonumber \\
   & & \qquad\qquad 
   + ~\widetilde{A}(k,q) \left(  g^{\mu\nu} \, - \,  \frac{k^\mu k^\nu}{k^2} \, - \, \frac{q^\mu q^\nu}{q^2}  \right)  
      \label{Sol:WTITwoPhotonTwoFermion2}
\end{eqnarray} 
where in last term the result of Eq. (\ref{EQ:L1}) weas used to rewrite $\lambda_1$.  In the approximation considered here, 
the longitudinal two-photon-two-fermion vertex $\Gamma^{\mu\nu}$ is completely determined by the fermion propagator functions $A$ and $B$.
Then, the propagator equations and the vertex equation define a closed set of equations that should be solved simultaneously.

Inserting  Eq. (\ref{Sol:WTITwoPhotonTwoFermion2}) into the vertex equation (\ref{Eq:DSE-vertex}) then, after some algebra,
it follows that the two-photon-two-fermion contribution to the vertex equation is
\begin{eqnarray}
& &
\MyIntPhoton ~ \Bigg\{
A_1 ~ B\big((p-q)^2\big)
      \Bigg[ k^{\mu } \, \frac{ (kq)^2}{\bar{k}^2 \bar{q}^2}   -   q^{\mu }  \, \frac{(kq)}{\bar{q}^2} \Bigg]
   \nonumber \\
   & & \hspace{2cm}
   ~ + ~ A_0 ~ B\big((p-q)^2\big) 
   \Bigg[   \frac{k^\mu}{k^2} \Bigg( - \, (kp)  + \frac{(kq) (pq)}{{q}^2} \Bigg) +\bar{p}^{\mu }  -  q^{\mu }  \frac{(pq)}{\bar{q}^2}  \Bigg]
   \nonumber \\
   & & \hspace{2cm}
  ~ + ~ 
  B_0 ~  A\big((p-q)^2\big)
        \Bigg[  \frac{k^\mu}{k^2} \Bigg(  - \, (kp)  + \frac{(kq) (pq)}{{q}^2} \Bigg)
                    + {p}^{\mu} - {q}^{\mu } \frac{ (pq)}{\bar{q}^2}
        \Bigg]
  ~ + ~ 3 \, A_3 ~ B\big((p-q)^2\big) ~ \frac{k^{\mu }}{{k}^2}
\nonumber \\
& & \qquad\quad
+ ~
A_1 ~  A\big((p-q)^2\big) \Bigg[
\frac{k^\mu}{k^2} \bigg( 
- \, \slashed{p} \, \frac{ (kq)^2}{{q}^2}  +   \slashed{k} \,\left( \frac{ (kq) \,  (pq)}{{q}^2}  - 2 \,  (kp) +(kq) \right)  +  \slashed{q} \, \frac{(kp) (kq) }{{q}^2} \Bigg)
   \nonumber \\
  & & \hspace{4.5cm}
    +  \frac{q^\mu}{q^2} \Bigg( (kq) \, \slashed{p} -  \slashed{k} (pq) -  (kp) \, \slashed{q} \Bigg)
    + {p}^{\mu } \, \slashed{k}
    + \gamma^\mu \Bigg(  (kp) - (kq) \Bigg) 
\Bigg]
\nonumber \\
& & \hspace{2cm}
 ~+ ~ 
  A_0 ~  A\big((p-q)^2\big) \Bigg[
  \frac{k^\mu}{k^2} \Bigg(  \slashed{k} \Big( (pq) - p^2 \Big)    + \slashed{q} \Big( \frac{p^2  (kq) }{q^2} -  (kp) \Big)  \Bigg)
                                          \nonumber \\
                                          & & \hspace{7cm}
                                          + ~ {p}^{\mu } \, \slashed{q} -  {q}^{\mu } \, \frac{{p}^2 }{{q}^2} \,  \slashed{q}
                                          + \gamma^\mu \Big( {p}^2 - \, (pq)    \Big)
 \Bigg]
 \nonumber \\
& & \hspace{2cm}
 ~ + ~ 
 A_2 ~  A\big((p-q)^2\big) \Bigg[
          \frac{k^\mu}{k^2} \Bigg(  \slashed{k} \Big( {q}^2  -   (pq) \Big) 
          +  \slashed{q} \Big( 2 \, \frac{(kq) (pq)}{\bar{q}^2}  -  (kp) - (kq) \Big) \Bigg)
          + p^{\mu } \, \slashed{q}
   \nonumber \\
   & & \hspace{7cm}
   + q^\mu \, \slashed{q}  \,  \Big( 1  - \, 2 \, \frac{(pq)}{\bar{q}^2} \Big) 
   + \, \gamma^\mu  \Big( (pq) - q^2 \Big)
 \Bigg]
 \nonumber \\
& & \hspace{2cm}
~ + ~ B_0 ~ B\big((p-q)^2\big) \Bigg[
\frac{k^\mu}{k^2} \Bigg( \slashed{q} \, \frac{(kq) }{{q}^2}  - \, \slashed{k} \Bigg)
                                   -\frac{\bar{q}^{\mu } \, \slashed{q}}{\bar{q}^2}+\, \gamma^\mu
\Bigg]
\nonumber \\
& & \hspace{2cm}
~ + ~
A_3 ~  A\big((p-q)^2\big) ~ \frac{k^\mu}{k^2} \Bigg[
\slashed{q} \, \Big( 3  - 2 \, \frac{ (pq) }{\bar{q}^2} \Big) -  \slashed{p} 
\Bigg]
\nonumber \\
& & \qquad\quad
~ + ~
i  ~  A_0 ~  A\big((p-q)^2\big)   ~  {\epsilon }^{\mu  \alpha \beta \eta} ~ {p}_\alpha {q}_\beta ~ \gamma_\eta \, \gamma_5
    \nonumber \\
    & & \hspace{2cm}
~ - ~ i ~  A_1 ~  A\big((p-q)^2\big) 
\Bigg[ 
 {\epsilon }^{\mu  \alpha \beta  \eta}  ~ p_\beta   -  {\epsilon }^{\mu  \alpha \beta \eta}  ~ q_\beta \Bigg] k_\alpha \gamma_\eta \, \gamma_5
    \nonumber \\
    & & \hspace{2cm}
~ +  ~ i ~ A_2 ~  A\big((p-q)^2\big)  ~ {\epsilon }^{\mu  \alpha \beta \eta} ~ p_\alpha q_\beta ~ \gamma_\eta \, \gamma_5
\nonumber \\
& & \qquad\quad
+ ~
A_0 ~ B\big((p-q)^2\big) 
\Bigg[ ~ \sigma^{\mu\alpha} p_\alpha 
          + \frac{k^\mu}{k^2} \, \sigma^{\alpha\beta} p_\alpha k_\beta 
          + \frac{1}{q^2} \, \Bigg( q^\mu - k^\mu \, \frac{(kq)}{k^2} \Bigg) \, \sigma^{\alpha\beta} p_\alpha q_\beta ~ 
\Bigg]
\nonumber \\
& & \hspace{1.5cm}
~ + ~  A_1 ~ B\big((p-q)^2\big)
 \Bigg[ ~ \sigma^{\mu\alpha} k_\alpha
              +  \frac{1}{q^2} \, \Bigg( q^\mu - k^\mu \, \frac{(kq)}{k^2} \Bigg) \sigma^{\alpha\beta} k_\alpha q_\beta ~
 \Bigg]
\nonumber \\
& & \hspace{1.5cm}
~ + ~ A_2 ~ B\big((p-q)^2\big) 
\Bigg[ ~  \sigma^{\mu\alpha} q_\alpha
                -  \frac{k^\mu}{k^2} \, \sigma^{\alpha\beta} k_\alpha q_\beta ~
\Bigg]
\nonumber \\
& & \hspace{1.5cm}
~ + ~ B_0 ~A\big((p-q)^2\big) 
  \Bigg[ ~ 
            \sigma^{\mu\alpha} p_\alpha
            -  \sigma^{\mu\alpha} q_\alpha
            + \frac{k^\mu}{k^2} \, \sigma^{\alpha\beta} p_\alpha k_\beta
            + \frac{k^\mu}{k^2} \, \sigma^{\alpha\beta} k_\alpha q_\beta
            \nonumber \\
            & & \hspace{7cm}
            + \frac{1}{q^2} \, \Bigg( q^\mu - k^\mu \frac{(kq)}{k^2} \Bigg) \sigma^{\alpha\beta} p_\alpha q_\beta
  \Bigg] ~
\Bigg\}
\label{EqRhSTwoPhoton}
\end{eqnarray}
where the various terms were grouped according to their tensor properties, where
\begin{equation}
\MyIntPhoton =  - \, i \, g^2 \, \int \frac{d^4q}{(2 \, \pi)^4} ~ \frac{ D(q^2) }{A^2\big( (p-q)^2 \big) \, (p-q)^2 - B^2\big( (p-q)^2 \big)}
\end{equation}
and
\begin{eqnarray}
   A_0 & = &  \Big( A\big(p^2\big) + A\big((p+k-q)^2\big) - A\big((p+k)^2\big) - A\big((p-q)^2\big)  \Big) / (kq) \ , \label{VertModel-A0} \\
   A_1 & = &  \Big( A\big((p+k-q)^2\big) - A\big((p+k)^2\big)   \Big) / (kq) \ , \label{VertModel-A1} \\
   A_2 & = &  \Big( A\big((p-q)^2\big) - A\big((p+k-q)^2\big)  \Big) / (kq) \ ,\label{VertModel-A2}  \\
   A_3 
   & = &  A\big(p^2\big) + A\big((p-q)^2\big) - A\big((p+k)^2\big) -  A\big((p+k-q)^2\big)   \\
      B_0 
   & = &  \Big(  B\big((p+k)^2\big) + B\big((p-q)^2\big) - B\big(p^2\big) - B\big((p+k-q)^2\big)  \Big) / (kq) \ , \  .
\end{eqnarray}
Once more, for $B = 0$ the above expression simplifies considerably and only the vector and axial vector terms give non-vanishing contributions.

The full r.h.s. of the vertex equation is given, taking the approximations mentioned, by the sum of Eqs (\ref{EqRhSVertex}), (\ref{EqRhSTwoPhoton})
and adding the tree level vertex.

\section{The on-shell vertex and its quiral limit \label{Sec:OnShell}}

Ket us now return to the analysis of the vertex equation and explore its tensorial decomposition in particular cases.
Let us start the analysis of photon-fermion vertex by looking the on-shell photon-fermion vertex $\widetilde{\Gamma}$ defined as
\begin{equation}
    \Big[ \bar u (p^\prime) ~  \widetilde{\Gamma}^\mu (p, \, -p^\prime = - p - k; \, k) ~  u(p) \Big]  = 
    \left. 
   \Big[ \bar u (p^\prime) ~  \Gamma^\mu (p, \, -p^\prime = - p - k; \, k) ~  u(p) \Big]
    \right|_{p^2=p^{\prime \, 2} = m^2} \ .
\end{equation}   
Assuming that the free Dirac equation can be applied to simplify the r.h.s of this equation, then, after some algebra, one arrives at
\begin{eqnarray}
\widetilde{\Gamma}^\mu (p, \, -p^\prime = - p - k; \, k)  & = &
\Big(p^\prime + p\Big)^\mu ~
    \Bigg\{
    - \, \lambda_3 + \Big( m^2 - (p^\prime p) \Big) \, \tau_1 + \tau_5  + 2 \, m \Big[ \lambda_2 + \Big( m^2 - (p^\prime p) \Big) \tau_2 \Big]
    \Bigg\}
    \nonumber \\
    & & \qquad
    + ~ \gamma^\mu ~ \Bigg\{
        \lambda_1 + 2 \, \Big( m^2 - (p^\prime p) \Big) \, \tau_3 - 2 \, m \, \tau_5
    \Bigg\}
    \nonumber \\
    & & \qquad\qquad
    + \, i ~~ {\tau_8}  ~~ {\gamma }_{\sigma } ~ {\gamma }_5 ~~ ~{\epsilon }^{\sigma  \mu  \alpha\beta} ~~  {p}_\alpha \, p^\prime_\beta
    \nonumber \\
    & = & 
    \gamma^\mu ~
    \Bigg\{
    \lambda_1 + 2 \, \Big( m^2 - (p^\prime p) \Big) \, \tau_3
             + 2 \, m \, \Bigg[
                              2 \, m \, \lambda_2  \, - \,  \lambda_3 + \Big( m^2 - (p^\prime p) \Big) \,  \Big( \tau_1 + 2 \, m \, \tau_2 \Big)
                              \Bigg]
    \Bigg\}
    \nonumber \\
    & & \qquad
    + ~
     \sigma^{\mu\alpha} k_\alpha ~ \Bigg\{
           - \, \lambda_3 + \Big( m^2 - (p^\prime p) \Big) \, \tau_1 + \tau_5  + 2 \, m \Big[ \lambda_2 + \Big( m^2 - (p^\prime p) \Big) \tau_2 \Big] 
     \Bigg\}
     \nonumber \\
     & & \qquad\qquad
         + \, i ~~ {\tau_8}  ~~ {\gamma }_{\sigma } ~ {\gamma }_5 ~~ ~{\epsilon }^{\sigma  \mu  \alpha\beta} ~~  {p}_\alpha \, p^\prime_\beta
     \label{OnShellVertex}
\end{eqnarray}
where the last expression was obtained from the first after using the Gordon identity
\begin{equation}
   2 \, m \, \bar u(p) \gamma^\mu u(p) = 
    \bar u(p) \bigg( \left( p^\prime + p \right)^\mu - \sigma^{\mu\alpha} \left( p^\prime - p \right)_\alpha  \bigg)u(p) \ .
\end{equation} 
The chiral on-shell photon-fermion vertex corresponds to take the limit $m \rightarrow 0$ and $\lambda_3 \rightarrow 0$  and reads
\begin{eqnarray}
\widetilde{\Gamma}^\mu (p, \, -p^\prime = - p - k; \, k)  & = &
    \gamma^\mu ~
    \Bigg\{
    \lambda_1 - 2 \,  (p^\prime p)  \, \tau_3
    \Bigg\}
~ + ~
     \sigma^{\mu\alpha} k_\alpha ~ \Bigg\{
             - (p^\prime p) \, \tau_1 + \tau_5  
     \Bigg\} 
~  + ~ i ~ {\tau_8}  ~~ {\gamma }_{\sigma } ~ {\gamma }_5 ~~ ~{\epsilon }^{\sigma  \mu  \alpha\beta} ~~  {p}_\alpha \, p^\prime_\beta  ,
  \label{OnShellVertexChiral}
\end{eqnarray}
i.e. it gets contributions only from $\lambda_1$, $\tau_1$, $\tau_3$ and $\tau_5$. It follows that the vertex models that do not consider 
these transverse form factors, as is the case of e.g. the Maris-Tandy or the Curtis-Pennington vertices, the chiral limit of the on-shell model 
is reduced to its tree level structure.

The anomalous magnetic and electric fermion form factors can be read from Eq. (\ref{OnShellVertex}) or, for their chiral limit, 
from Eq. (\ref{OnShellVertexChiral}).
Note that the vertex Ward-Takahashi identity for the photon-fermion vertex implies
\begin{eqnarray}
k_\mu ~ \Big[ \bar u (p^\prime) ~  \widetilde{\Gamma}^\mu (p, \, -p^\prime = - p - k; \, k) ~  u(p) \Big] & = &
k_\mu ~ \Big[ \bar u (p^\prime) ~  {\Gamma}^\mu (p, \, -p^\prime = - p - k; \, k) ~  u(p) \Big] 
\nonumber \\
& = &  \Big[ \bar u (p^\prime) ~  \bigg( S^{-1}(-p^\prime) -  S^{-1}(p) \bigg) ~  u(p) \Big] 
\nonumber \\
& = &  \Big[ \bar u (p^\prime) ~  \bigg( B(p^2) - B(p^{\prime \, 2})  - m  \big( A(p^{\prime \, 2}) + A(p^2) \big)  \bigg) ~  u(p) \Big] 
\end{eqnarray}
that vanishes either in the chiral limit or when $A$ and $B$ are momentum independent functions, as occurs for the tree solution of QED.
The chiral limit of the of the photon-fermion vertex will be discussed further in Sec . \ref{Sec:ChiralVertex}.

The results derived in this section are quite general and are also valid for QCD, after the correction of the color structure and after taking into
consideration the contribution associated with $\Lambda_4$. 
If for QED the on-shell condition seems to be a reasonable approximation, in QCD the on-shell condition is conceptually difficult 
to accepted due to confinement. Moreover, there are important differences between the longitudinal form factors for QED and QCD.
The Ward-Takahashi identity for the vertex in QED is replaced, in QCD, by a Slavnov-Taylor identity (STI)
which takes into account the quark-ghost scattering kernel and whose solution for the $\lambda_i$ is considerable more
complex when compared to the Abelian vertex.
For example, the general solution of the STI in QCD for the longitudinal form factors gives a non-vanishing $\lambda_4$.
However, in lowest order in the coupling constant, the perturbative solution of the Slavnov-Taylor identity for the vertex reproduces the results 
for the longitudinal vertex of the Abelian theory and, in this sense, it is tempting to use the Ball-Chiu vertex in QCD, despite its know limitations.
A discussion on the longitudinal form factors in QCD can be found in e.g.
\cite{Oliveira:2018fkj,Alkofer:2000wg,Fischer:2006ub,RichardWilliams2007,Aguilar:2010cn,Aguilar:2016lbe,Binosi:2016wcx,Aguilar:2018epe,Aguilar:2018csq,Oliveira:2018ukh,Oliveira:2020yac} and references therein.
See also \cite{Mena:2023mqj} for a recent discussion of the QCD corrections to the on-shell photon-quark vertex.

\section{The photon-fermion Vertex in the soft photon limit \label{Sec:softphotonlimit}}

The vertex soft photon limit  is defined for zero photon momentum. For this kinematics the expression for the vertex DSE becomes
\begin{eqnarray}
   {\Gamma}^\mu (p, \, -p ; \, 0)   & = &
   \, 2 \, \lambda_3 (p^2, \, p^2, 0)  ~ p^\mu ~ + ~    \lambda_1 (p^2, \, p^2, 0)  ~  {\gamma }^{\mu } ~ +  ~  4 \, \lambda_2 (p^2, \, p^2, 0) 
  ~ \slashed{p} \, p^\mu  
\nonumber \\
&  = & - \, 2 \, B^\prime (p^2)  ~ p^\mu ~ + ~    A(p^2)  ~  {\gamma }^{\mu } ~ +  ~  2 \, A^\prime (p^2)  ~ \slashed{p} \, p^\mu \ .
\nonumber \\
& = &   
    \gamma^\mu  ~ + ~ i \, g^2 \, \int \frac{d^4q}{(2 \, \pi)^4} ~D_{\zeta\zeta^\prime}(q)  ~ \gamma^\zeta ~ S(p-q)
   \nonumber \\
& & 
\qquad\qquad
   \Bigg\{ ~ 
                                 \, {\Gamma}^\mu ( p - q, \, -p + q; \, 0) \,
                                S(p-q) \, {\Gamma}^{\zeta^\prime} (p -q, \, -p; q) 
~ + ~
                                 \, {\Gamma}^{\zeta^\prime\mu} (p - q, \, -p; \, q , \, 0) 
                                \Bigg\}  \ .
                                \label{Eq:DSE-vertex-soft}
\end{eqnarray}
The two vertices ${\Gamma}^\mu (p, \, -p ; \, 0)$ and $ {\Gamma}^\mu ( p - q, \, -p + q; \, 0)$ appearing in the r.h.s. of above equation
require only the longitudinal form factors, that are determined by their vertex WTI. 
Indeed, looking at Eq. (\ref{FullVertex-Dirac}), if there are no kinematical singularities associated with
the transverse form factors $\tau_i$, then the decomposition of the l.h.s. of the vertex equation calls only for the $\lambda_i$.
On the other hand, the longitudinal component of the two-photon-two-fermion vertex for the kinematics appearing in the above equation
was obtained by solving the corresponding WTI in \cite{Oliveira:2022bar}. The solution of this WTI gives $\Gamma^{\mu\nu}$
in terms of the fermion propagator functions $A$ and $B$ and their derivatives; see the solution given in \cite{Oliveira:2022bar} that is reproduced in
Eq. (\ref{Eq:Gamma-mu-nu-q0}).
In general, the two-photon-two-fermion vertex $\Gamma^{\mu\nu}$ is given by longitudinal and transverse, relative
to the photons momenta, components and, therefore, Eq. (\ref{Eq:DSE-vertex-soft}) can provide a consistency check on the various contributions
of  $\Gamma^{\mu\nu}$. In practice this consistency check is difficult to implement as the equation still involves a full vertex
contribution that is ${\Gamma}^{\zeta^\prime} (p -q, \, -p; q)$.

For completeness, we provide the expression for the two-photon-two-fermion derived in \cite{Oliveira:2022bar} that solves the associated WTI
when one of the photon momentum vanishes
\begin{eqnarray}
\Gamma^{\zeta\mu} (p-q, \, -p; \, q, \, 0) & = &
  2 ~ g^{\zeta\mu} ~ B^\prime(p^2) ~ + ~  2 ~ \frac{q^\zeta p^\mu - q^\zeta q^\mu}{q^2}  ~ \Bigg[ B^\prime (p^2) - B^\prime\Big((p-q)^2\Big) \Bigg]
  \nonumber \\
  & & \quad
  ~ + ~ \frac{q^\zeta \gamma^\mu}{q^2} ~ \Bigg[  A\Big((p-q)^2\Big) - A (p^2)   \Bigg]
  \nonumber \\
  & & \quad\quad
  ~ - ~ 2 ~ \slashed{q} ~ \frac{q^\zeta p^\mu - q^\zeta q^\mu}{q^2} ~  A^\prime\Big((p-q)^2\Big) 
  \nonumber \\
  & & \quad\quad\quad
    ~ - ~ 2 ~ \slashed{p} \Bigg\{ ~ g^{\zeta\mu} ~A^\prime(p^2) ~ + ~ \frac{q^\zeta p^\mu - q^\zeta q^\mu}{q^2}  ~ \Bigg[ A^\prime (p^2) - A^\prime\Big((p-q)^2\Big) \Bigg] ~ \Bigg\} \ ,
    \label{Eq:Gamma-mu-nu-q0}
\end{eqnarray}
where the solutions of the vertex WTI for the various $\lambda_i$ was used.

\section{The Chiral Vertex \label{Sec:ChiralVertex}}

In the chiral limit and in the absence of dynamical chiral symmetry breaking not only $B = 0$ but also $\lambda_3 = 0$ and $B_0 = A_3 = 0$ 
in the two-photon-two-fermion contribution to the r.h.s. of the vertex equation (\ref{Eq:DSE-vertex}).
By ignoring the contribution of the transverse part of the vertex, i.e. assuming that
\begin{equation}
  \Gamma_\mu = \lambda_1 \, L^{(1)}_\mu + \lambda_2 \, L^{(2)}_\mu  \ ,
  \label{ChiralVertex-0}
\end{equation}  
a straightforward calculation shows that, for any general linear covariant gauge,
there are no contributions for the scalar, pseudoscalar and tensor components of (\ref{FullVertex-Dirac}) and, therefore,
\begin{equation}
    \tau_{1, \, 4, \, 5, \, 7} = 0 \ ,
    \label{Constraints-OnShellVertex-ChiralLimit}
\end{equation}
in good agreement with the analysis of the perturbative one-loop QED calculation for the vertex in a general linear covariant gauge of \cite{Kizilersu:1995iz},
that is a special case of that just considered. 
Note that by using the vertex as given in Eq. (\ref{ChiralVertex-0}) in the r.h.s. of the vertex equation, the analysis of Eq. (\ref{Eq:DSE-vertex}) 
goes beyond the traditional first order in the coupling constant perturbative solution that is recovered for $\lambda_1 = 1$, $\lambda_2 = 0$ and  setting
$\Gamma^{\mu\nu} = 0$. Indeed if $\lambda_2 = 0$ in the analsys of Eq. (\ref{Eq:DSE-vertex}), 
the description of the vertex becomes closer to the perturbative solution. It follows that, in first order in the coupling constant,
and for the vertex model discussed in Sec. \ref{Sec:rhsVertex}, the r.h.s. of the vertex equation simplifies and reads
\begin{eqnarray}
& & 
   {\Gamma}^\mu (p, \, -p -k; \, k)   =  \gamma^\mu  
\nonumber \\
& &
   ~ - ~ i \, g^2 \, \int \frac{d^4q}{(2 \, \pi)^4} ~ D(q^2)  ~ \frac{\lambda_1 \Big((p +k - q)^2, (p - q)^2, k^2 \Big)}{A^2 ((p-q)^2) (p-q)^2 - B^2 ((p-q)^2)} 
   ~ \frac{\lambda_1\Big( (p+k)^2, (p +k-q)^2, q^2)  \Big) }{A^2 ((p+k-q)^2) (p+k-q)^2 - B^2 ((p+k-q)^2)}
\nonumber \\
& & \qquad
\Bigg\{    
     k^{\mu } \left( \slashed{q} - 3 \, \slashed{p} \right)
     ~ +  ~  3 \,  \gamma^{\mu } \Bigg[ (kp) - (kq) - 2 \,  (pq) + p^2 + q^2 \Bigg]
     ~ - ~ 3 \,  \Big( p^{\mu } - q^{\mu } \Big) \Big( \slashed{k} + 2 \, \slashed{p} \Big)
     \nonumber \\
     & & \qquad\qquad
     + ~ 2 \, \frac{\slashed{q} }{\bar{q}^2}
          \Bigg[ k^{\mu } (pq) + p^{\mu } \Big( (kq) + 2 \, (pq) + q^2 \Big) - q^{\mu } \Big( (kp) + p^2 + 2 \, q^2\Big)
          \Bigg]
     \nonumber \\
     & & \qquad\qquad
    + ~ i ~ \Bigg[
               \epsilon^{\mu \alpha\beta\eta} ~ k_\alpha p_\beta 
               + \left(2 \, \frac{(pq)}{q^2} - 3 \right) ~  \epsilon^{\mu \alpha\beta\eta} ~ k_\alpha q_\beta 
               -  2 \, \frac{(kq)}{q^2} ~  \epsilon^{\mu \alpha\beta\eta} ~ p_\alpha q_\beta
         \Bigg] ~\gamma_\eta \gamma_5
       \Bigg\}
\nonumber \\
& &
   ~ - ~ i \, g^2 \, \int \frac{d^4q}{(2 \, \pi)^4} ~  D(q^2) 
               ~ \frac{\lambda_1 \Big((p +k - q)^2, (p - q)^2, k^2 \Big)}{A^2 ((p-q)^2) (p-q)^2 - B^2 ((p-q)^2)} 
\nonumber \\
& & \qquad
   \Bigg\{ 
   A_0 \Bigg[
                 \left( \frac{(pq)}{{k}^2}  - \frac{ p^2}{k^2} \right) \slashed{k}  \, {k}^{\mu }
                 + \frac{(kq)}{k^2} \slashed{p} \,  k^{\mu }
                 - \frac{(kp)}{\bar{k}^2} \slashed{q}\,  {k}^{\mu }
                 + \Big( {p}^2   -  (pq) \Big) \gamma^{\mu }
                 - \slashed{p}  \, q^{\mu } 
                 + \slashed{q}  \, p^{\mu } 
            \Bigg]
\nonumber \\
& & \qquad\qquad            
  + ~ A_1 \Bigg[
  2 \,  \left( \frac{(kq)}{{k}^2} -  \frac{ (kp)}{{k}^2} \right) \slashed{k} \, {k}^{\mu }
  +  \Big( (kp) - (kq) \Big) \gamma^\mu
  + \slashed{k} \, p^{\mu }  
  -  \slashed{k} \, q^{\mu } 
                  \Bigg]
\nonumber \\
& & \qquad\qquad            
    + ~ A_2 \Bigg[
    \left( \frac{{q}^2}{k^2}  - \frac{ (pq)}{{k}^2} \right) \slashed{k} \, {k}^{\mu }
    + \frac{(kq)}{{k}^2} \slashed{p}  \, {k}^{\mu }
    - \frac{(kp)}{{k}^2} \slashed{q}  \, {k}^{\mu }
    + \Big( (pq)  - {q}^2 \Big) \gamma^\mu
    - \slashed{p} \, {q}^{\mu }  
    + \slashed{q}  \, {p}^{\mu } 
    \Bigg]
\nonumber \\
& & \qquad\qquad
-  ~ i ~ \Bigg[
 - \, A_0 ~ \epsilon^{\mu \alpha\beta\eta} ~ p_\alpha q_\beta
+ A_1 \Bigg( \epsilon^{\mu \alpha\beta\eta} ~ k_\alpha p_\beta  -  \epsilon^{\mu \alpha\beta\eta} ~ k_\alpha q_\beta \Bigg)
- \,  A_2 \, \epsilon^{\mu \alpha\beta\eta} ~ p_\alpha q_\beta
    \Bigg] ~\gamma_\eta\gamma_5
                     \Bigg\} 
\nonumber \\
& &
   ~ + ~ i \, g^2 \, \int \frac{d^4q}{(2 \, \pi)^4} ~  \left( D(q^2) - \frac{\xi}{q^2} \right)
               ~ \frac{\lambda_1 \Big((p +k - q)^2, (p - q)^2, k^2 \Big)}{A^2 ((p-q)^2) (p-q)^2 - B^2 ((p-q)^2)} 
\nonumber \\
& & \qquad                                 
   \Bigg\{ 
    A_0 \Bigg[
            \frac{{p}^2 (kq)}{{k}^2 {q}^2} \slashed{q} \, {k}^{\mu }
            - \frac{(kq)}{{k}^2} \slashed{p} \, {k}^{\mu }
            + \slashed{p} \, {q}^{\mu }
            -\frac{ {p}^2}{{q}^2} \slashed{q} \, {q}^{\mu } 
            \Bigg]
\nonumber \\
& & \qquad \qquad                                
  +  A_1 \Bigg[
   \left( \frac{(kq) (pq)}{{k}^2 {q}^2}    - \frac{ (kq) }{{k}^2} \right) \slashed{k} \, {k}^{\mu }
  + \frac{(kp) (kq)}{{k}^2 {q}^2} \slashed{q} \, {k}^{\mu }
  -   \frac{(kq)^2 }{{k}^2 \bar{q}^2} \slashed{p} \, {k}^{\mu } 
  + \left( 1   - \frac{ (pq)}{{q}^2} \right) \slashed{k}  \, {q}^{\mu } 
  + \frac{(kq)}{{q}^2} \slashed{p} \, {q}^{\mu } 
  - \frac{ (kp)}{{q}^2}   \slashed{q}  \, {q}^{\mu }
  \Bigg]
\nonumber \\
& & \qquad   \qquad                              
    + A_2 \Bigg[
    \left( 2 \, \frac{ (kq) (pq)}{{k}^2 {q}^2}  - \frac{(kq)}{\bar{k}^2} \right) \slashed{q} \, {k}^{\mu }
    - \frac{(kq)}{{k}^2} \slashed{p} \, {k}^{\mu }
    + \slashed{p}  \, {q}^{\mu } 
    + \left( 1 - 2 \, \frac{ (pq)}{\bar{q}^2}  \right) \slashed{q} \, {q}^{\mu }
    \Bigg]
\Bigg\} \ 
\label{Eq:DSE-vertex-chiral}
\end{eqnarray}
where
\begin{equation}
\lambda_1 \Big((p +k - q)^2, (p - q)^2, k^2 \Big) ~ = ~ \frac{1}{2} \Bigg( A \Big( (p - q)^2\Big) ~ + ~ A \Big( (p +k - q)^2\Big)\Bigg)
\end{equation}
and
\begin{equation}
 \lambda_1\Big( (p+k)^2, (p +k-q)^2, q^2)  \Big) ~ = ~ \frac{1}{2} \Bigg( A \Big( (p +k - q)^2\Big) ~ + ~ A \Big( (p +k )^2\Big)\Bigg) \ . 
\end{equation}
are the solutions of the  WTI for the vertex.

The constraints derived for the chiral limit, see Eq. (\ref{Constraints-OnShellVertex-ChiralLimit}), simplify further 
the on-shell vertex in the massless limit (\ref{OnShellVertexChiral}) that reduces to 
\begin{eqnarray}
\widetilde{\Gamma}^\mu (p, \, -p^\prime = - p - k; \, k)  & = &
 \gamma^\mu \bigg[ \lambda_1 - 2 \, (p^\prime p) \, \tau_3   \bigg]
  ~ + ~ i ~ \tau_8 ~ \gamma_\sigma\gamma_5 ~ \epsilon^{\sigma\mu\alpha\beta} ~ p_\alpha \, p^\prime_\beta 
  ~ + ~ \sigma^{\mu\alpha}\,k_\alpha ~ \tau_5  \ .
  \label{OnShellVertexChiral-withContraints}
\end{eqnarray}
and from its definition it follows that
\begin{eqnarray}
 k_\mu \Big[ \bar u(p^\prime) ~ \widetilde{\Gamma}^\mu (p, \, -p^\prime = - p - k; \, k)  ~ u(p) \Big]  = 0 \ ,
\end{eqnarray}
recovering the usual textbook expression for the current conservation.
Note that to arrive at this later result, the role of the free particle Dirac equation and, therefore, of the on-shell concept is fundamental.

\section{Transverse Form Factors from the Photon-Fermion vertex DSE \label{SecTransverseFF}}
 
Let us return to the discussion on the computation of the transverse form factor $\tau_i$ from the photon-fermion vertex DSE (\ref{Eq:DSE-vertex})
using  its tensor decomposition as in Eq. (\ref{FullVertex-Dirac}). 
The expressions derived from this last equation, see below, are exact and it is only when the vertex model discussed previously to simplify the writing of
the r.h.s of the vertex equation is used that they become approximate. 
Moreover, the results given for each of the transverse form factors $\tau_i$, they explore the tensor decomposition of the vertex in the tensor basis
and they are not only exact but also be easily translated to QCD, that requires the introduction of the correction coming  the color degrees of freedom and
a non-vanishing $\lambda_4$.

\subsection{The form factor $\tau_1$}

In Minkowski spacetime the transverse form factor $\tau_1$ can be computed using Eq. (\ref{tau1-fromvertex}) and it is given by
the solution of equation
\begin{equation}
  \tau_1  \Big((p+k)^2, \, p^2, \, k^2\Big)=   ~ \frac{1}{ p^2 \, k^2 - (pk)^2 } \Bigg\{
\frac{1}{4} \, \text{Tr} \bigg( p_\mu \,{\Gamma}^\mu (p, \, -p -k; \, k)   \bigg) ~ + ~ \frac{ 2 \, p^2 + (pk) }{(p+k)^2 \, - \, p^2} ~ \Bigg( B((p+k)^2) - B( p^2) \Bigg) 
\Bigg\} \ .
\label{tau1-fromvertex-Mink}
\end{equation}   
It is instructive to compare the outcome of this last expression with the computation of the same form factor in perturbation theory at
one-loop perturbative for QED \cite{Kizilersu:1995iz}. It can be observed that the overall factor $1/( p^2 \, k^2 - (pk)^2)$ appears
in the perturbative and in the exact description of vertex equation. Moreover, this overall factor is considered in some vertex
models used to study gauge theories, see  
\cite{Albino:2021rvj,El-Bennich:2022obe,Lessa:2022wqc} and references therein,
but not all vertex models, as e.g. in \cite{Bashir:2011dp}. Moreover, in
some cases $\tau_1$ is not considered at all \cite{Curtis:1990zs,Curtis:1992jg,Kizilersu:2009kg,Aguilar:2012rz,Oliveira:2020yac}.
Note also that for the tree level solution, where $B$ is momentum independent, this form factor vanish.

Inserting in Eq. (\ref{tau1-fromvertex-Mink}) the r.h.s of the DSE vertex equation using the approximation to the vertex 
discussed in Sec. \ref{Sec:rhsVertex} it turns out that $\tau_1$ is proportional to momentum integrals that require the fermion propagator $B$.
This result is the equivalent to the one-loop perturbation QED calculation \cite{Kizilersu:1995iz} that gives a form factor
$\tau_1$ proportional to the mass of the fermion. 
Then, for massless fermions the prediction of perturbation theory and that of the vertex model in the Landau gauge considered for r.h.s. of the vertex
equation is $\tau_1 = 0$.

The expression for the form factor $\tau_1$ is a scalar equation and it is possible to write it in Euclidean spacetime with the rules given in 
App. \ref{AppWick}. Indeed,  for this transverse form factor, the Euclidean spacetime solution reads
\begin{equation}
  \left. \tau_1  \Big((p+k)^2, \, p^2, \, k^2\Big) \right|_E =    \frac{1}{ p^2 \, k^2 - (pk)^2 } \Bigg\{
  \left. \frac{1}{4} \, \text{Tr} \bigg( p_\mu \,{\Gamma}^\mu (p, \, -p -k; \, k)   \bigg)  \right|_E  +  \frac{ 2 \, p^2 + (pk) }{k^2 + 2 (pk) } ~ \Bigg( B((p+k)^2) - B( p^2) \Bigg) 
\Bigg\}
\label{tau1-fromvertex-Eucl}
\end{equation} 
where the scalar coming from the trace should be Wick rotated after building the scalar itself, i.e. after taking the trace in Minkowski spacetime.

\subsection{The form factor $\tau_8$}

The transverse form factor $\tau_8$ can be read from the axial term appearing in Eq. (\ref{FullVertex-Dirac}). In this case is harder to provide an
exact equation to compute $\tau_8$. However, inserting in the decomposition of the r.h.s. of the vertex equation the vertex model approximation 
discussed previously, it turns out that
\begin{eqnarray}
&&
i ~~ {\tau_8}  \Big((p+k)^2, \, p^2, \, k^2\Big)  ~~ ~{\epsilon }^{\eta  \mu  \alpha\beta} ~~  {p}_\alpha \, {k}_\beta ~
= 
\nonumber \\
& & \hspace{1cm}\quad
 = ~ - ~  \, \frac{g^2}{4} \, \int \frac{d^4 q}{(2 \, \pi)^4} ~ 
      \frac{D(q^2)}{A \Big( (p-q)^2 \Big) \,  A \Big( (p+k-q)^2 \Big)}  
      \nonumber \\
      & & \hspace{4.5cm}
       \frac{1}{ (p-q)^2 ~ - ~  M^2 \Big( (p-q)^2 \Big) } ~
       \frac{1}{(p+k-q)^2 ~ - ~  M^2 \Big( (p+k-q)^2 \Big) } 
 \nonumber \\
 & & \hspace{4.5cm} 
    \Bigg[ A \Big( (p+k-q)^2 \Big) + A \Big( (p-q)^2 \Big) \Bigg] \Bigg[ A \Big( (p+k-q)^2 \Big) + A \Big( (p+k)^2 \Big) \Bigg]
 \nonumber \\
  & & \hspace{4.5cm}
     \Bigg\{  \Bigg( 2 \,   \frac{(pq)}{{q}^2}  + 1 \Bigg) ~  \epsilon^{\mu \alpha \beta \eta} ~ k_\alpha \, q_\beta 
    ~ - ~ 2 \,  
    \frac{ (kq)}{{q}^2}  ~ \epsilon ^{\mu \alpha \beta \eta} ~ p_\alpha q_\beta
    ~ - ~ 3  ~  \epsilon^{\mu \alpha \beta \eta} ~ k_\alpha p_\beta 
    \Bigg\} 
 \nonumber \\
 & & \hspace{2cm}
  + ~ g^2 \, \int \frac{d^4q}{(2 \, \pi)^4} ~ \frac{ D(q^2) }{A\big( (p-q)^2 \big)}
  \frac{ 1 }{ (p-q)^2 - M^2\big( (p-q)^2 \big)}
  \nonumber \\
  & & \hspace{4.5cm}
  \Bigg\{
  ~  \Bigg( A_0 ~+~ A_2 \Bigg)     ~  {\epsilon }^{\mu  \alpha \beta \eta} ~ {p}_\alpha {q}_\beta 
~ + ~   A_1 ~~  {\epsilon }^{\mu  \alpha \beta  \eta}  ~ p_\alpha \, k_\beta   
~ + ~ A_1 ~~  {\epsilon }^{\mu  \alpha \beta \eta}  ~ q_\beta  k_\alpha 
\Bigg\}
     \label{FF-tau8-0}
\end{eqnarray}
where $M = B/A$ is the running fermion mass. From this last equation a scalar can be built by multiplication with 
${\epsilon }_{\eta  \mu  \alpha\beta} ~  {p}^\alpha \, {k}^\beta$ and gives
\begin{eqnarray}
&&
 {\tau_8}  \Big((p+k)^2, \, p^2, \, k^2\Big)  \Big( p^2 k^2 - (pk)^2 \Big) ~
= 
\nonumber \\
& & \quad
 = ~ i ~  \, \frac{g^2}{4} \, \int \frac{d^4 q}{(2 \, \pi)^4} ~ 
      \frac{D(q^2)}{A \Big( (p-q)^2 \Big) \,  A \Big( (p+k-q)^2 \Big)}  
      \nonumber \\
      & & \hspace{3.5cm}
       \frac{1}{ (p-q)^2 ~ - ~  M^2 \Big( (p-q)^2 \Big) } ~
       \frac{1}{(p+k-q)^2 ~ - ~  M^2 \Big( (p+k-q)^2 \Big) } 
 \nonumber \\
 & & \hspace{3.5cm} 
    \Bigg[ A \Big( (p+k-q)^2 \Big) + A \Big( (p-q)^2 \Big) \Bigg] \Bigg[ A \Big( (p+k-q)^2 \Big) + A \Big( (p+k)^2 \Big) \Bigg]
 \nonumber \\
  & & \hspace{2cm}
     \Bigg\{   \Bigg( 2 \,   \frac{(pq)}{{q}^2}  + 1 \Bigg) ~ \Big(  k^2 (pq) - (pk) (kq)   \Big)
    ~ + ~ 2 \,  
    \frac{ (kq)}{{q}^2}  \Big( p^2 (kq) - (pq) (pk) \Big)
    ~ + ~ 3  \Big( (pk)^2 - p^2k^2\Big) 
    \Bigg\} 
 \nonumber \\
 & &
  + ~ i ~ g^2 \, \int \frac{d^4q}{(2 \, \pi)^4} ~ \frac{ D(q^2) }{A\big( (p-q)^2 \big)}
  \frac{ 1 }{ (p-q)^2 - M^2\big( (p-q)^2 \big)}
  \nonumber \\
  & & \hspace{2cm}
  \Bigg\{
   \Bigg( A_0 ~+~ A_2 \Bigg)    \Big( p^2 (kq) - (pq) (pk) )\Big) 
~ + ~   A_1 \Big( p^2 k^2 - (pk)^2  +  (pq) k^2 - (pk) (kq) \Big)
\Bigg\}
     \label{FF-tau8}
\end{eqnarray}
The quantities $A_0$, $A_1$ and $A_2$ come from the contribution of the two-photon-two-fermion vertex. There expressions are given
in Eqs (\ref{VertModel-A0}), (\ref{VertModel-A1}) and (\ref{VertModel-A2}), respectively, and are sums of differences of the propagator
form factor $A$ evaluated at different momenta. As already discussed, for a tree level vertex it comes that $\tau_8 = 0$.
The contractions of the above expressions involving the Levi-Civita tensor are examples of the type of structures that can contribute
to $\tau_8$. Moreover, the various Levi-Civita terms considered and are the only type of structures that can contribute to this form factor.

As happens for $\tau_1$, the form factor $\tau_8$ has the same overall multiplicative factor $1/(p^2k^2 - (pk)^2)$  that 
is also observed in the one-loop perturbative calculation but not in some of the usual vertex models. Indeed, some of the models
write $\tau_8$ proportional to $1/( (p+k)^2 - p^2 ) = 1/(p^2 + k^2 + 2 \, (pk) )$.

The expression for $\tau_8$ in Eq. (\ref{FF-tau8}) is a scalar function and, therefore, after performing the Wick rotation one arrives
at its Euclidean spacetime version that is
\begin{eqnarray}
&&
 {\tau_8}  \Big((p+k)^2, \, p^2, \, k^2\Big)   ~
 = ~   \, \frac{g^2}{4} \, \int \frac{d^4 q}{(2 \, \pi)^4} ~ 
      \frac{D(q^2)}{A \Big( (p-q)^2 \Big) \,  A \Big( (p+k-q)^2 \Big)}  
      \nonumber \\
      & & \hspace{5cm}
       \frac{1}{ (p-q)^2 ~ + ~  M^2 \Big( (p-q)^2 \Big) } ~
       \frac{1}{(p+k-q)^2 ~ + ~  M^2 \Big( (p+k-q)^2 \Big) } 
 \nonumber \\
 & & \hspace{5cm} 
    \Bigg[ A \Big( (p+k-q)^2 \Big) + A \Big( (p-q)^2 \Big) \Bigg] \Bigg[ A \Big( (p+k-q)^2 \Big) + A \Big( (p+k)^2 \Big) \Bigg]
 \nonumber \\
  & & \hspace{5cm}
     \Bigg\{   \Bigg( 2 \,   \frac{(pq)}{{q}^2}  + 1 \Bigg) ~ \frac{ k^2 (pq) - (pk) (kq) }{p^2 k^2 - (pk)^2}
    ~ + ~ 2 \,     \frac{ (kq)}{{q}^2}  \frac{ p^2 (kq) - (pq) (pk) }{p^2 k^2 - (pk)^2}
    ~ - ~ 3  
    \Bigg\} 
 \nonumber \\
 & & \hspace{3.5cm}
  + ~  g^2 \, \int \frac{d^4q}{(2 \, \pi)^4} ~ \frac{ D(q^2) }{A\big( (p-q)^2 \big)}
  \frac{ 1 }{ (p-q)^2 + M^2\big( (p-q)^2 \big)}
  \nonumber \\
  & & \hspace{5cm}
  \Bigg\{
   \Bigg( A_0 ~+~ A_2 \Bigg)  \frac{ p^2 (kq) - (pq) (pk) ) }{p^2 k^2 - (pk)^2}
~ + ~   A_1 \Bigg( 1  +  \frac{(pq) k^2 - (pk) (kq)}{p^2 k^2 - (pk)^2}\Bigg) 
\Bigg\}  \ .
     \label{FF-tau8-Eucl}
\end{eqnarray}
If $A$ is independent of the momentum, then further simplifications occur and
\begin{eqnarray}
&&
 {\tau_8}  \Big((p+k)^2, \, p^2, \, k^2\Big)   ~
 = ~   \, g^2 \, \int \frac{d^4 q}{(2 \, \pi)^4} ~ 
      D(q^2) ~
       \frac{1}{ (p-q)^2 ~ + ~  M^2 \Big( (p-q)^2 \Big) } ~
       \frac{1}{(p+k-q)^2 ~ + ~  M^2 \Big( (p+k-q)^2 \Big) } 
 \nonumber \\
  & & \hspace{5cm}
     \Bigg\{   \Bigg( 2 \,   \frac{(pq)}{{q}^2}  + 1 \Bigg) ~ \frac{ k^2 (pq) - (pk) (kq) }{p^2 k^2 - (pk)^2}
    ~ + ~ 2 \,     \frac{ (kq)}{{q}^2}  \frac{ p^2 (kq) - (pq) (pk) }{p^2 k^2 - (pk)^2}
    ~ - ~ 3  
    \Bigg\}  \ .
     \label{FF-tau8-Eucl-A-constante}
\end{eqnarray}

\subsection{The form factors $\tau_2$, $\tau_3$ and $\tau_6$}

The transverse form factors $\tau_2$, $\tau_3$ and $\tau_6$
can be computed from Eq. (\ref{FullVertex-Dirac})  combined with (\ref{FullVertex-Dirac-Contraction}) by looking at their vectorial part. 
Indeed, it follows that
\begin{eqnarray}
& & 
\gamma_\mu \, {\Gamma}^\mu (p, \, -p -k; \, k)   ~ = ~ 
   \lambda_1 \, d_D
   ~ + ~ \lambda_2 \Bigg[ k^2  ~ + ~ 4 \, (pk) ~ + ~ 4 \,  p^2\Bigg]
   ~ + ~ 2 \, \tau_2 \Bigg[  p^2 \, k^2 ~ - ~   (pk)^2  \Bigg]
   ~ + ~ \tau_3 \Bigg[ k^2 (d_D-1) \Bigg]
   \nonumber \\
   & & \hspace{3cm}
   + ~ \tau_6 \Bigg[  (d_D - 1) \Bigg(  k^2  \, + \,  2 \,  (pk) \Bigg)  \Bigg]
     \label{Eq:ContTau236-1}
\end{eqnarray}
where we give the result after computing $1/4$ of the trace of the l.h.s. of the equation and where $d_D$ is the number of spacetime dimensions. 
To access the transverse form factors let us consider also
\begin{eqnarray}
& & 
\slashed{k} ~ \Bigg( p_\mu \,{\Gamma}^\mu (p, \, -p -k; \, k)   \Bigg) ~ = ~ 
   \lambda_1 \, (pk)
   ~ + ~ \lambda_2 \Bigg[  \bigg( k^2 + 2 \, (pk) \bigg) \,  \bigg( (pk) + 2 \, p^2 \bigg)   \Bigg]
   \nonumber \\
   & & \hspace{3cm}
   ~ + ~ \tau_2 \Bigg[ \bigg( k^2 +  2 \, (pk) \bigg)\, \bigg( p^2 k^2 - (pk)^2 \bigg) \Bigg]
   ~ + ~ 2 \,  \tau_6 \Bigg[  (pk)^2  -   p^2 k^2  \Bigg]
      \label{Eq:ContTau236-2}
\end{eqnarray}
and
\begin{eqnarray}
& &
\slashed{p} ~ \Bigg(  p_\mu \,{\Gamma}^\mu (p, \, -p -k; \, k) \Bigg)  ~ = ~ 
  \lambda_1 \, p^2
  ~ + ~ \lambda_2 \Bigg[ \bigg( (pk) + 2 \, p^2 \bigg)^2 \Bigg]
  ~ + ~ \tau_2 \Bigg[ \bigg( (pk)  + 2 \, p^2 \bigg) \bigg( p^2 k^2  - (pk)^2 \bigg)\Bigg]
   \nonumber \\
   & & \hspace{3cm}
   + ~ \tau_3 \Bigg[ p^2 k^2 -  (pk)^2\Bigg]
   ~ + ~ \tau_6 \Bigg[ p^2 k^2  -  (pk)^2   \Bigg] 
        \label{Eq:ContTau236-3}
\end{eqnarray}
after taking the trace of the various expressions (omitted to simplify the notation).
Then, combining Eqs (\ref{Eq:ContTau236-1}) to (\ref{Eq:ContTau236-3}) it is possible to disentangle $\tau_2$, $\tau_3$ and $\tau_8$.
In App. \ref{Apptau236} the details for getting $\tau_2$, $\tau_3$ and $\tau_8$ from the above expressions are reported. We call
the reader attention, once more, that the formal expressions (\ref{GetTau2}) to (\ref{GetTau6}) have no approximations and
come from the decomposition of the l.h.s. of the vertex DSE. As seen in  App. \ref{Apptau236}, these three form factors
all share the common factor 
\begin{equation}
  \varphi = - \, \frac{1}{8} \, \frac{1}{p^2 \, k^2}  \, \frac{1}{ p^2 k^2 - (pk)^2 }
\end{equation}  
that is only partially accommodated in some of the usual vertex models.
Further, as discussed in App. \ref{Apptau236}, for a tree level
vertex the form factors $\tau_2$, $\tau_3$ and $\tau_6$ they all vanish.

\subsection{The form factors $\tau_4$, $\tau_5$ and $\tau_7$}

The remaining transverse form factors $\tau_4$, $\tau_5$ and $\tau_7$ can be computed by combining the tensor components of the 
decomposition of the l.h.s. of the vertex DSE.
The calculation, that results in an exact equation, is discussed in App. \ref{Apptau457}, where the interested reader can find
explicit and exact expressions for each of the form factors. Note that these form factors have in common an overall kinematical
factor and , once more, for a tree level vertex $\tau_4 = \tau_5 = \tau_7 = 0$.

The results for the form factors derived from the tensor terms are valid only for QED. However, the modification require to generalize the 
results to QCD are relative simple and come mainly from having $\lambda_4 \ne 0$ in QCD, 
while in QED the solution of the vertex WTI returns a vanishing $\lambda_4$.

\section{Summary and Conclusions \label{Sec:Summary}}

In this work the Dyson-Schwinger equation for the photon-fermion vertex was investigated with the help of a tensor basis that provides a full description of
this vertex. Exact expressions that enable the computation of all the vertex transverse form factors were derived based only on the vertex DSE. 
These expressions, that depend only on the tensorial basis, are quite general and can be applied both to the QED photon-fermion vertex and to 
the QCD quark-gluon vertex, after taking into account the color structure and correcting for the contribution of $\lambda_4$. 
The solutions for the transverse form factors use the information coming from the vertex Ward-Takahashi identity that determines its four
longitudinal form factors. 

The relations derived are independent of the dynamics of the theory, that is not the same for QED or QCD. 
The dynamics  comes in when, going beyond the tensorial decomposition of the vertex, one attempt to describe
the integral part of the equation.
However, exploring only the tensorial basis for the vertex allows to understand the relative importance of the contribution of the various
form factors to the propagator equations and, therefore, to chiral symmetry breaking, and also enable a discussion of the IR divergences in QED.
Moreover, it allows to address the photon-fermion vertex in certain limits and, in particular, its on-shell version, its chiral limit and its soft photon limit.
Our approach re-derived some of the results found for the one-loop vertex calculation in a more general framework.
From the expressions derived for the vertex in particular cases, exact results that take into account all QED form factors were obtained 
for the anomalous magnetic and anomalous electric fermion couplings. 

The solutions for the various transverse QED form factors suggests that the parametrization of these form factors should include certain kinematical
factors, that are not always taken into account in the vertex models. The perturbative solution for the vertex is included in our 
analysis, as a special case, and we hope that the study performed is robust and complies with multiplicative renormalizability, that is not investigated here.

Along the way to obtain the results described, we discussed a simplified vertex model for the photon-fermion QED vertex that goes beyond
the perturbative solution of the vertex and that enables the computation of all the transverse form factors  in Minkowsky spacetime or
in Euclidean spacetime.  Indeed, the vertex model to feed the r.h.s. of the vertex equation considers only a tree level type of vertex and
includes a contribution due to the two-photon-two-fermion one-particle irreducible Green function. 
The vertex model does not comply fully with gauge invariance, as it solves approximately the vertex Ward-Takahashi 
identity. However, the vertex model can be improved at the expenses of handling larger expressions.

Herein, none of the transverse form factors is computed explicitly, despite arriving at closed expressions for each of the $\tau_i$. The computation 
of the transverse form factor using the vertex model or its generalizations  will be considered in a follow up publication. 

\section*{Acknowledgements}

This work was partly supported by the FCT – Funda\c{c}\~ao para a Ci\^encia e a Tecnologia, I.P., under Projects Nos. UIDB/04564/2020, UIDP/04564/2020.
The author also acknowledges financial support from  grant 2022/05328-3, from São Paulo Research Foundation (FAPESP).
The author thanks T. Frederico and W. de Paula for helpful discussions.

\begin{appendix}

\section{Disentangling  $\tau_2$, $\tau_3$ and $\tau_6$ \label{Apptau236}}

The transverse form factors $\tau_2$, $\tau_3$ and $\tau_6$ can be disentangle solving the linear systems
of equations (\ref{Eq:ContTau236-1}) to (\ref{Eq:ContTau236-3}). To simplify the writing of the form factors, let us introduce the notation
\begin{eqnarray}
  \Delta & = & p^2 \, k^2 ~ - ~ (pk)^2 \ ,  \label{Eq:DELTAFF} \\
  \delta & = & k^2 ~ + ~ 2 \, (pk) \ , \\
  \zeta & = & 2 \, p^2 ~ + ~ (pk) 
\end{eqnarray}
and
\begin{eqnarray}
\Lambda_0 & = & \gamma_\mu \, {\Gamma}^\mu (p, \, -p -k; \, k)   ~ - ~
   \lambda_1 \, d_D  ~ - ~ \lambda_2 \Bigg[ k^2  ~ + ~ 4 \, (pk) ~ + ~ 4 \,  p^2\Bigg] \ , \\
\Lambda_1 & = & \slashed{k} ~ \Bigg( p_\mu \,{\Gamma}^\mu (p, \, -p -k; \, k)   \Bigg) ~ - ~ 
   \lambda_1 \, (pk)  ~ - ~ \lambda_2 \Bigg[  \bigg( k^2 + 2 \, (pk) \bigg) \,  \bigg( (pk) + 2 \, p^2 \bigg)   \Bigg] \ , \\
\Lambda_2 & = & \slashed{p} ~ \Bigg(  p_\mu \,{\Gamma}^\mu (p, \, -p -k; \, k) \Bigg)  ~ - ~ 
  \lambda_1 \, p^2  ~ - ~ \lambda_2 \Bigg[ \bigg( (pk) + 2 \, p^2 \bigg)^2 \Bigg] \ .
\end{eqnarray}
It follows that 
\begin{eqnarray}
 \tau_2 \Big( (p +k)^2, \, p^2, k^2 \Big)  & = & \varphi  ~ \Bigg\{ 2 ~  \Delta ~ \Lambda_0  + \big( d_D - 1 \big) \Big(  \delta   -  k^2  \Big) ~ \Lambda_1  + 2 \, k^2 \,  \Big( 1 -  d_D \Big) ~ \Lambda_2\Bigg\} \ , 
  \label{GetTau2}. \\
 \tau_3 \Big( (p +k)^2, \, p^2, k^2 \Big)& = & \varphi ~ \Bigg\{  - \, \Big(  \delta   + 2 \,  \zeta \Big) \,  \Delta ~ \Lambda_0 + \Bigg(  \delta \,  \zeta \Big( 1 -  d_D \Big) +2 \, \Delta \Bigg) ~ \Lambda_1 
              ~+ ~ \Bigg(  \delta ^2 \Big( d_D - 1 \Big)+ 4 \, \Delta  \Bigg) ~\Lambda_2 \Bigg\} \ , 
   \label{GetTau3} \\
 \tau_6 \Big( (p +k)^2, \, p^2, k^2 \Big) & = & \varphi ~ \Bigg\{  \delta  \, \Delta ~ \Lambda_0 ~ + ~ \Bigg(  \Big( d_D - 1 \Big)  \zeta \,  k^2  - 2 \, \Delta  \Bigg) ~ \Lambda_1 ~ + ~ \delta  \, k^2 \, \Big( 1 -   d_D  \Big) ~ \Lambda_2 \Bigg\}
 \label{GetTau6}
\end{eqnarray}
with
\begin{eqnarray}
\varphi = \frac{1}
       {\Delta  \Bigg( 4 \, \Delta ~ + ~  \Big( d_D - 1 \Big) \Big( \delta ^2   ~ - ~ 2 \,  \zeta  \, k^2  ~ - ~ \delta \, k^2  \Big)  \Bigg)} \ .
\end{eqnarray}
For the tree level vertex it follows that $\Lambda_0 = \Lambda_1 = \Lambda_2 = 0$ and, therefore, all three form factors 
$\tau_2$, $\tau_3$ and $\tau_6 = 0$ vanish.

\section{Disentangling $\tau_4$, $\tau_5$ and $\tau_7$ \label{Apptau457}}

In order to compute the transverse form factors from vertex equation (\ref{FullVertex-Dirac}), let consider only its tensorial component given by
\begin{eqnarray}
  T^\mu & = & 
     \frac{1}{2} \, \sigma^{\mu\alpha} \, k_\alpha \,  \Bigg[ 2 \, \tau_5 \, + \, \tau_7 \Big( k^2 + 2 \, (kp) \Big) \Bigg]
     \nonumber \\
     & & \qquad\qquad
     + \,  
          \sigma^{\mu\alpha} \, p_\alpha \,  \Bigg[ \,  \tau_7 \Big( k^2 + 2 \, (kp) \Big) \, \Bigg]
     \nonumber \\
     & & \qquad\qquad\qquad\qquad
     \, + \,  \sigma^{\alpha\beta} \, p_\alpha \, k_\beta \Bigg[ \, k^\mu \, \Big( \tau_7 -  (pk) \, \tau_4  \Big) +  p^\mu \Big(  k^2 \, \tau_4  +  2 \, \tau_7  \Big) \Bigg]
     \nonumber
\end{eqnarray}
and defined the following contractions
\begin{equation}
T_0 = \frac{1}{4} \mbox{Tr } \Bigg( \sigma^{\mu\alpha} k_\alpha \, T_\mu \Bigg) \, , \qquad
T_1 = \frac{1}{4} \mbox{Tr } \Bigg( \sigma^{\mu\alpha} p_\alpha \, T_\mu \Bigg) \, , \qquad\mbox{ and }\qquad
T_0 = \frac{1}{4} \mbox{Tr } \Bigg( \sigma^{\alpha\beta} p_\alpha k_\beta \, T_\mu \Bigg) p^\mu \ .
\end{equation}
then it comes that
\begin{eqnarray}
& & 
\tau_4 \Big( (p +k)^2, \, p^2, k^2 \Big) = 
\frac{ \Delta  \, T_0 
         + 2 \, \Delta  \, T_1
         - 3  \, T_2 \Big( 2 \, (pk) + k^2 \Big)
         }
  {2 \Delta ^2 \Big( 2 \, (pk) + k^2 \Big)}
  \\   & &
  \tau_5 \Big( (p +k)^2, \, p^2, k^2 \Big) =
  \frac{ T_0 \, \Big(  (pk) + 2 \, p^2 \Big)
           -  T_1 \, \Big( 2  \, (pk) + k^2 \Big)
           - 2 \, T_2  }
  {- \, 4 \,  \Delta}
   \\   & & 
   \tau_7 \Big( (p +k)^2, \, p^2, k^2 \Big) = 
  \frac{ \, T_0  \, (pk) 
            - \, T_1 \, k^2}  {2 \, \Big( 2 \, (pk) +  k^2 \Big) \, \Delta}
\end{eqnarray}
where $\Delta$ is defined in Eq. (\ref{Eq:DELTAFF}). As before, the integral expressions appearing in the r.h.s. of the vertex equation can be read from
Eqs (\ref{EqRhSVertex}) and (\ref{EqRhSTwoPhoton}).

\section{Wick Rotation to Euclidean Spacetime \label{AppWick}}

The formal manipulation discussed in the current work are performed in Minkowski spacetime. However, the solution of the DSE becomes
numerically easier in Euclidean spacetime. Indeed, doing the Wick rotation to Euclidean spacetime possible singularities and other structures that make the Minkowski numerical solutions difficult are avoided.  The Euclidean equations are obtained from the Minkowski equations
by a naive Wick rotation that are performed applying the following rules
\begin{equation}
  p^2 \rightarrow - \, p^2_E \ , \qquad
  A(p^2) \rightarrow A_E(- \, p^2_E) \ , \qquad
  B(p^2) \rightarrow B_E(- \, p^2_E) \ , \qquad
  D(p^2) \rightarrow - D_E(- \, p^2_E) 
\end{equation}  
together with the replacement
\begin{equation}
  \int d^4p  \rightarrow i  \,   \int d^4p_E \ .
\end{equation}  
In these expressions the Euclidean quantities are named with the subscript $E$. 

\end{appendix}
 

\end{document}